%% file: DGMM2025_arXiv.tex
\documentclass[runningheads]{llncs}
\usepackage[T1]{fontenc}
\usepackage{graphicx}
\usepackage{hyperref}
\usepackage[dvipsnames]{xcolor}

\usepackage{amsmath}
\usepackage{amssymb}
\usepackage{tikz}
\usepackage{wrapfig}
\usepackage{environ}
\usepackage{bbm}

\usepackage{changes}
\usepackage{algorithm}
\usepackage[noend]{algpseudocode}
\usepackage{varwidth}
\usepackage{mathtools}
\usepackage{pgfplots}
\usepackage[utf8]{inputenc}
\pgfplotsset{compat=1.18}
\usetikzlibrary {spy,arrows.meta,decorations.shapes}
\setauthormarkup{}
\definechangesauthor[name=jo, color=blue]{JO}
\definechangesauthor[name=rom, color=teal]{ROM}

\input{./macros}

\colorlet{MyGreen}{green!80!gray}

\begin{document}
    \title{Fast and exact visibility on digitized shapes and application to saliency-aware normal estimation}
    \titlerunning{Fast and exact visibility on digitized shapes}
    \author{Romain Negro\inst{1}\orcidID{0009-0000-5866-4991} \and
    Jacques-Olivier Lachaud\inst{1}\orcidID{0000-0003-4236-2133}}
    \authorrunning{R. Negro and J.-O. Lachaud}
    \institute{Universit\'e Savoie Mont Blanc, CNRS, LAMA, F-73000 Chambéry, France\\
    \email{\{romain.negro|jacques-olivier.lachaud\}@univ-smb.fr}}
    \maketitle              %
    \begin{abstract}
        Computing visibility on a geometric object requires heavy
        computations since it requires to identify pairs of points that
        are visible to each other, i.e.\ there is a straight segment
        joining them that stays in the close vicinity of the object
        boundary. We propose to exploit a specific representation of
        digital sets based on lists of integral intervals in order to
        compute efficiently the complete visibility graph between
        lattice points of the digital shape. As a quite direct
        application, we show then how we can use visibility to estimate
        the normal vector field of a digital shape in an accurate and
        convergent manner while staying aware of the salient and sharp features of
        the shape.

    \end{abstract}

    \section{Introduction}
    \input{introduction}

    \section{Visibility through tangency of chords}
    \input{visibility}

    \section{Fast computation using integer intervals intersections}
    \input{computation}

    \section{Saliency-aware normal and curvature estimation}
    \input{normalsandcurvature}

    \section{Conclusion}
    \input{conclusion}

    \begin{credits}
        \subsubsection{\ackname}
        This work is partially supported by the French National Research Agency
        within the StableProxies project (ANR-22-CE46-0006).
    \end{credits}
    \bibliographystyle{splncs04}
    \bibliography{biblio}

\end{document}

%% file: macros.tex
\newcommand{\R}{\ensuremath{\mathbb{R}}}
\newcommand{\Z}{\ensuremath{\mathbb{Z}}}
\let\C\relax \DeclareMathOperator{\C}{\ensuremath{\mathcal{C}^d}} %

\newcommand{\ve}[0]{\ensuremath{\mathbf{e}}}

\newcommand{\vv}[0]{\ensuremath{\mathbf{v}}}

\newcommand{\Star}[1]{\ensuremath{\mathrm{Star}\left(#1\right)}}

\newcommand{\Le}{\ensuremath{\leqslant}}
\newcommand{\Ge}{\ensuremath{\geqslant}}

%% file: introduction.tex
    Visibility is a fundamental concept in computational geometry and
    digital topology, with applications ranging from computer vision
    (occlusions), geometric modeling (feature detection,
    geodesics) to computer graphics (path tracing,
    shadowmap). Visibility within polygons in the plane has been
    extensively studied in the litterature~\cite{ghosh:2007-book} and
    is still studied in higher dimensional
    spaces~\cite{orourke:2017-book}. The general problem has a high
    complexity. For instance, the 3d visibility complex between $n$
    spheres and occluding spheres has a complexity
    $O(n^4)$~\cite{durand:2002-tog}. Exact 3d visibility between
    polygons and occluding polygons is often cast into 5d Plücker
    space~\cite{nirenstein:2002-ewr}, which nicely represents 3d lines
    and their respective positions. The computational cost remains
    high even with decomposition into coarse cells for speed up
    (several days of computation for 1M triangles at the time).

    To increase efficiency, several authors have cast the
    visibility problem into a digital space, generally $\Z^3$ or its
    decomposition into cubical cells. Visibility is simplified as
    whether there is a connected digital straight line
    joining a pair of cells without occluding cell(s)
    (e.g.\ Soille~\cite{soille:1994-prl} uses Bresenham
    segments). Coeurjolly \emph{et al.}~\cite{coeurjolly:2004-prl}
    use a non-symmetric visibility definition and preimage
    computations to speed up these
    algorithms. Chica~\cite{chica:2008-spm} uses a half-cell erosion
    of the complementary space and Bresenham lines to decide
    visibility.

    We propose an algorithm that solves the following visibility
    problem: given a digital set $X \subset \Z^d$, two points $p,q$ of
    $X$ are \emph{visible} whenever the Euclidean straight line
    segment $\lbrack p,q \rbrack$ is never at chessboard distance
    ($\infty$-distance) greater or equal to $1$ from $X$. It is
    equivalent to the \emph{cotangency} of $p$ and $q$ in $X$, as
    formalized in~\cite{lachaud:2021-dgmm,lachaud:2022-jmiv}, and can
    be viewed as an inclusion problem between sets of lattice points
    in this form. Indeed, cells surounding $X$ can be encoded as
    lattice points (Khalimsky coding), as well as the cells covering
    lattice directions. The originality of our method is to encode
    subsets of lattice points as sets of integer intervals and to
    reduce all the inclusion problems as intersections and
    translations of integer intervals. For a given lattice vector
    $\mathbf{t}$, we solve the visibility of every pair of points
    $(p,p+\mathbf{t})$ of $X$. Global visibility is achieved by
    testing all sought directions within a given range, and is
    parallelized straightforwardly.

    We focus on solving efficiently this problem because it is crucial
    in computing geometric normals along digital surfaces that are
    aware of salient features. Indeed, 2d discrete tangent estimators
    based on tangency or discrete line segments have proven to be both
    sensitive to sharp corners and multigrid convergent on the
    boundary of digitized smooth shapes~\cite{feschet:1999-dgci,lachaud:2007-ivc,nguyen:2011-pr}.
    Many ad hoc methods have been proposed for tangent plane or normal
    estimation on 3d digital surfaces
    (e.g. \cite{fourey:2009-cg,charrier:2011-iwcia,Cuel:2014-dgci,Lachaud:2017-lnm,mareche:2024-ispr}).
    If two of these methods~\cite{Cuel:2014-dgci,Lachaud:2017-lnm}
    have been shown to be multigrid convergent on digitization of
    shapes with smooth boundary, their formulation intrinsically
    smoothes features, e.g.\ in contrast with the method of
    Mar{ê}ch{é} \emph{et al.}~\cite{mareche:2024-ispr}, whose
    convergence is not established. We propose here a normal estimator
    defined as the most orthogonal vector to the cone of visibility at
    the point of interest. Visibility is high in smooth regions and
    the induced visibility vectors are good approximations of tangent
    vectors. On the contrary the visibility is blocked at sharp
    features, so the normal vector is not influenced by the
    geometry on the other side of the features. We demonstrate the
    superiority of this normal estimator for estimating curvatures
    along digital surfaces with sharp features.

    Note that our algorithm for computing visibility is in fact much
    more general. The same algorithm could be used to recognize if any
    pattern of lattice points is present on a shape. It is similar to
    an erosion algorithm where the structuring element is the chosen
    pattern, and could also be used for morphological operations~\cite{soille1999morphological}.
    The paper outline is as follows. In Section~2, we recall some
    definitions and present tangency of chords and its equivalence
    to visibility. Section~3 describes the specific data structure
    for encoding set of grid cells, and the main algorithm that
    exploits this encoding to compute visibility exactly. Section~4
    presents experimental results on visibility, its usage as a
    discrete normal estimator, and how it improves curvature estimates
    on digital surfaces. Finally, Section 5 concludes and gives some
    perspectives to this work.

%% file: visibility.tex
    All the theory and algorithms presented in this section are
    defined and valid in arbitrary dimension $d$. They will however be
    illustrated in 2d for clarity, while experiments will be
    performed on 3d digital shapes.  Let $\Z^d$ be the $d$-dimensional
    digital space.  Let $\mathcal{C}^d$ be the (cubical) cell complex
    induced by the lattice $\Z^d$ : its 0-cells are the points of
    $\Z^d$, its 1-cells are the open unit segments joining two 0-cells
    at distance 1, its 2-cells are the open unit squares, \ldots, and
    its $d$-cells are the $d$-dimensional open unit hypercubes with
    vertices in $\Z^d$.  We denote by $\mathcal{C}^d_k$ the set of its
    $k$-cells, for $0 \Le k \Le d$.  In the following, a cell always
    designates an element of $\mathcal{C}^d$, and the term subcomplex
    always indicates a subset of $\mathcal{C}^d$.

    The \emph{topological closure} of a cell $\tau$ is denoted by
    $\bar{\tau}$. The \emph{star} of a cell $\sigma$ is the subcomplex
    $\Star{\sigma}\coloneqq\{\tau \in \mathcal{C}, \sigma \subset \bar{\tau}
    \}$. It is naturally extended to any subcomplex $K$ by union,
    i.e. $\Star{K} \coloneqq \cup_{\sigma \in K} \Star{\sigma}$, and more
    generally to any subset $Y$ of $\R^d$ as $\Star{Y}\coloneqq\{ c \in \C,
    \bar{c} \cap Y \neq \emptyset \}$. The \emph{body} $\|K\|$ of $K$
    is its realization in $\R^d$, i.e. $\|K\|\coloneqq\cup_{c \in K} c$. One
    can check that $\Star{K}=\Star{\|K\|}$, so these definitions are
    sound.

\begin{figure}[t]
      \centering
      \input{visibility.tikz}
      \caption{Examples of visibility and non visibility in 2D within
        the set $X$ (represented with black dots $\bullet$).}
      \label{fig:visibility-2d}
    \end{figure}
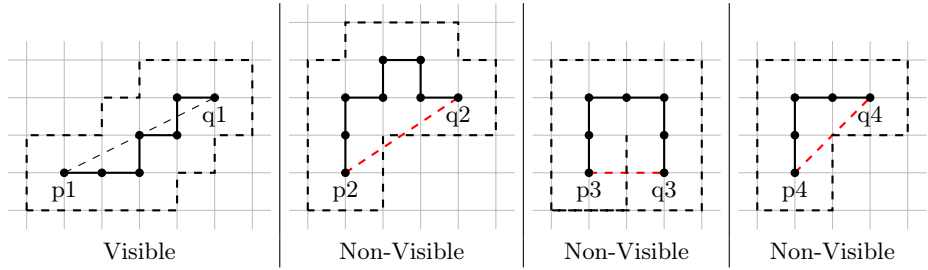

    \begin{definition}[Visibility]
      Let $Z \subset \Z^d$ be a digital set. Two digital points $p$
      and $q$ are \emph{visible in $Z$} if and only if the straight
      segment $[p, q]$ is included in $\Star{Z}$, i.e., $\Star{[p, q]}
      \subseteq \Star{Z}$.
    \end{definition}
    This definition corresponds to the concept of tangency presented
    in~\cite{lachaud:2022-jmiv}, limited to pairs of points. One can
    check that $p$ and $q$ must belong to $Z$ to be visible.
    Figure~\ref{fig:visibility-2d} illustrates what is considered
    visible or not on some examples.

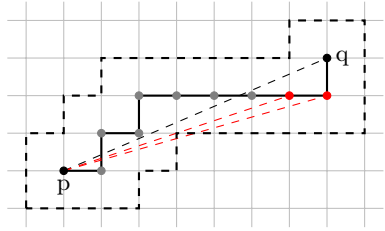
\begin{figure}[t]
        \centering
        \begin{tikzpicture}
            \draw[step=0.5,lightgray,thin,xshift=-1cm,yshift=-1cm] (0.25,0.25) grid (5.25,3.25);
            \draw[thick] (0,0) -- (0.5,0) -- (0.5,0.5) -- (1,0.5) -- (1,1) -- (3.5,1) -- (3.5,1.5);
            \draw [black, dashed] (0,0) -- (3.5,1.5);
            \draw [red, dashed] (0,0) -- (3.5,1);
            \draw [red, dashed] (0,0) -- (3,1);
            \filldraw[black] (0,0) circle (0.05) node[anchor=north] {p};
            \filldraw[gray] (0.5,0) circle (0.05);
            \filldraw[gray] (0.5,0.5) circle (0.05);
            \filldraw[gray] (1,0.5) circle (0.05);
            \filldraw[gray] (1,1) circle (0.05);
            \filldraw[gray] (1.5,1) circle (0.05);
            \filldraw[gray] (2,1) circle (0.05);
            \filldraw[gray] (2.5,1) circle (0.05);
            \filldraw[red] (3,1) circle (0.05);
            \filldraw[red] (3.5,1) circle (0.05);
            \filldraw[black] (3.5,1.5) circle (0.05) node[anchor=west] {q};
            \draw[thick,dashed] (-0.5,-0.5) -- (-0.5,0.5) -- (0,0.5) -- (0,1) -- (0.5,1) -- (0.5,1.5) -- (3,1.5) -- (3,2) -- (4,2) -- (4,0.5) -- (1.5,0.5) -- (1.5,0) -- (1,0) -- (1,-0.5) -- (-0.5,-0.5);
        \end{tikzpicture}
        \caption{The set of visible points in $X$ from source $p$ is not $1$-connected.}
        \label{fig:visibility-2d-not-connected}
    \end{figure}

    A quite unexpected fact about visibility it that the set of
    visible points from a source $p$ is not necessarily digitally
    connected in $X$. Indeed, for $n \in \Z, n \Ge 1$, we say that
    $p,q$ are \emph{$n$-neighbors} when $\|q-p\|_\infty \Le n$, and
    this induces the $n$-adjacency and $n$-connectedness relation in
    $X$. As illustrated on
    Figure~\ref{fig:visibility-2d-not-connected}, the set of points
    visible from $p$ is clearly not $1$-connected. As our experiments
    have shown, we even conjecture that it may not be $n$-connected
    for $n$ arbitrarily large.

    This characteristic has implications for the applicability of~\cite[Algorithm 3]{lachaud:2022-jmiv}, which is a breadth-first
    like algorithm for computing all the points of $X$ visible in $X$
    to a source point $p$. This algorithm sometimes outputs an
    incomplete collection of visible points. This is why in the next
    section we present an algorithm that computes the exact visibility
    pairs within $X$, without recourse to connectedness in its formulation.

%% file: visibility.tikz
\begin{tabular}{c@{~}|@{~}c@{~}|@{~}c@{~}|@{~}c}
  \begin{tikzpicture}
    \draw[step=0.5,lightgray,thin,xshift=-1cm,yshift=-1cm] (0.25,0.25) grid (3.75,2.75);
    \draw[dashed] (0,0) -- (2,1);
    \filldraw[black] (0,0) circle (0.05) node[anchor=north] {p1};
    \filldraw[black] (2,1) circle (0.05) node[anchor=north] {q1};
    \foreach \x/\y in {1/0,2/0,2/1,3/1,3/2} {
      \filldraw[black] (0.5*\x,0.5*\y) circle (0.05);
    }
    \draw [thick] (0,0) -- (0.5,0) -- (1,0) -- (1,0.5) -- (1.5,0.5) -- (1.5,1) -- (2,1);
    \draw[thick,dashed] (-0.5,-0.5) -- (-0.5,0.5) -- (0.5,0.5) -- (0.5,1) -- (1,1) -- (1,1.5) -- (2.5,1.5) -- (2.5,0.5) -- (2,0.5) -- (2,0) -- (1.5,0) -- (1.5,-0.5) -- (-0.5,-0.5);
  \end{tikzpicture} &
  \begin{tikzpicture}
    \draw[step=0.5,lightgray,thin,xshift=-1cm,yshift=-1cm] (0.25,0.25) grid (3.25,3.25);
    \draw[red,dashed,thick] (0,0) -- (1.5,1);
    \filldraw[black] (0,0) circle (0.05) node[anchor=north] {p2};
    \filldraw[black] (1.5,1) circle (0.05) node[anchor=north] {q2};
    \foreach \x/\y in {0/1,0/2,1/2,1/3,2/3,2/2} {
      \filldraw[black] (0.5*\x,0.5*\y) circle (0.05);
    }
    \draw [thick] (0,0) -- (0,1) -- (0.5,1) -- (0.5,1.5) -- (1,1.5) -- (1,1) -- (1.5,1);
    \draw[thick,dashed] (-0.5,-0.5) -- (-0.5,1.5) -- (0,1.5) -- (0,2) -- (1.5,2) -- (1.5,1.5) -- (2,1.5)  -- (2,0.5) -- (0.5,0.5) -- (0.5,-0.5) -- (-0.5,-0.5);
  \end{tikzpicture} &
  \begin{tikzpicture}
    \draw[step=0.5,lightgray,thin,xshift=-1cm,yshift=-1cm] (0.25,0.25) grid (2.75,2.75);
    \draw[red,dashed,thick] (0,0) -- (1,0);
    \filldraw[black] (0,0) circle (0.05) node[anchor=north] {p3};
    \filldraw[black] (1,0) circle (0.05) node[anchor=north] {q3};
    \foreach \x/\y in {0/1,0/2,1/2,2/2,2/1} {
      \filldraw[black] (0.5*\x,0.5*\y) circle (0.05);
    }
    \draw [thick] (0,0) -- (0,1) -- (1,1) -- (1,0);
    \draw[thick,dashed] (-0.5,-0.5) -- (-0.5,1.5) -- (1.5,1.5) -- (1.5,-0.5) -- (-0.5,-0.5) -- (0.5,-0.5) -- (0.5,0.5);
  \end{tikzpicture} &
  \begin{tikzpicture}
    \draw[step=0.5,lightgray,thin,xshift=-1cm,yshift=-1cm] (0.25,0.25) grid (2.75,2.75);
    \draw[red,dashed,thick] (0,0) -- (1,1);
    \filldraw[black] (0,0) circle (0.05) node[anchor=north] {p4};
    \filldraw[black] (1,1) circle (0.05) node[anchor=north] {q4};
    \foreach \x/\y in {0/1,0/2,1/2} {
      \filldraw[black] (0.5*\x,0.5*\y) circle (0.05);
    }
    \draw [thick] (0,0) -- (0,1) -- (1,1);
    \draw[thick,dashed] (-0.5,-0.5) -- (-0.5,1.5) -- (1.5,1.5) -- (1.5,0.5) -- (0.5,0.5) -- (0.5,-0.5) -- (-0.5,-0.5);
  \end{tikzpicture} \\
  Visible     & Non-Visible                   &
  Non-Visible & Non-Visible
\end{tabular}

%% file: computation.tex
As said in the introduction, any cell of $\mathcal{C}$ can be
characterized as a point with integer coordinates, aka \emph{lattice
point}. For instance, a standard way is to double the coordinates of
the centroid of the cell (often called its Khalimsky code). Now we
have to represent an arbitrary set $K$ of lattice points.

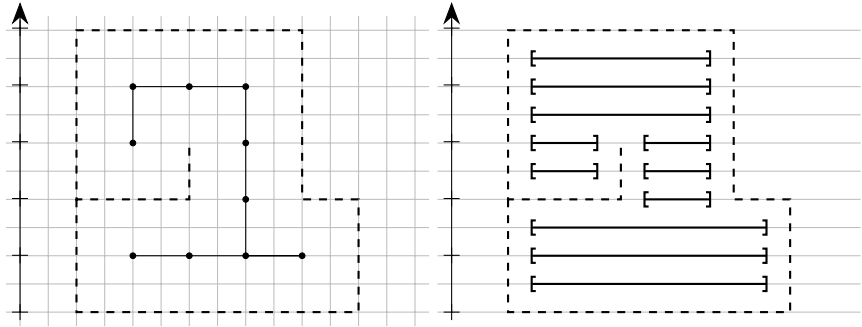
\begin{figure}[t]
  \centering
  \input{lattice.tikz}
  \caption{\label{fig:lattice-representation} Representation of a
    2d cell complex (here the star of a given curve) as a lattice
    map. From 63 cells (with two coordinates) on the left, we get 11
    intervals on the right.}
\end{figure}

Let $(\ve_j)_{j=1,\dots,d}$ be the canonical basis of $\Z^d$.  We may
choose an arbitrary projection \emph{axis} $j \in
\{1,\ldots,d\}$. Letting $\pi_j$ be the projector along this axis, we
can collect all the cells of $K$ that projects onto the same
points. Let $p \in K$. Then $\pi^{-1}_j(p) \cap \Star{K}$ is a set of cells having
all the same coordinates except along coordinate $j$. This set can be
ordered increasingly and stored as a list of integer intervals (see
Figure~\ref{fig:lattice-representation} for an illustration).

A \emph{sequence of intervals} is an ordered list of intervals $L =
([a_i,b_i])_{i=1,\ldots,n}$ such that $a_i \in \Z, b_i \in \Z, b_i + 1
< a_{i+1}$, i.e., disjoint intervals with at least a missing
integer. We denote by $\mathbb{L}$ the set of sequences of
intervals. An integer $k$ belongs to $L$ iff there exists $i$ such that
$a_i \Le k \Le b_i$. For a finite set of integers, there is a unique
sequence of intervals representing it.

A \emph{lattice map along axis $j$} is a set of pairs $(q,L_q)$, with
$q \in \Z^{d-1}$ and $L_q \in \mathbb{L}$. The lattice map $M_K$
\emph{represents} $K$ when, for any point $p \in K$, the pair
$(\pi_j(p), L_{\pi_j(p)})$ exists in $M_K$ and $p_j \in L_{\pi_j(p)}$,
with $p_j$ the $j$-th coordinate of $p$, and reciprocally, $\forall
(q,L_q) \in M_K, \forall i \in L_q, q+i\ve_j \in K$. A lattice map is
thus made of pairs $(q,L_q)$, where $q$ is the \emph{shift} while
$L_q$ is the \emph{intervals}. For a point $q$ which is a shift of
$M_K$, we write $M_K[q]$ for the corresponding intervals $L_q$.

The global algorithm for computing visibility is given in
Algorithm~\ref{alg:visibility}.  Given a digital input surface $C$ (a
subcomplex), we compute the lattice map $\Omega$ of $\Star{C}$ once,
generally along the axis where $C$ is the most elongated. In order to
compute visibility up to a given distance $r$, we consider all non-null
primitive vectors $D$ with infinity norm no greater than $r$.
For each vector $\vv$, we compute the lattice map $M_\vv$ of its
intersected cells (as a vector from $\mathbf{0}$ to $\vv$ in
$\R^d$). The visibility along the direction $\vv$ reduces to computing
for each pair of $M_\vv$ the possible translations of intervals in
$\Omega$, and then intersecting all the translations.

\begin{algorithm}
  \caption{Given a subcomplex $C$ and an integer $r$, returns the visibility from every point of $C$ up to distance $r$. The main axis is supposed to be $z$, while $x,y$ are the auxiliary axes.}
  \label{alg:visibility}
  \begin{algorithmic}
    \Function{Visibility}{$C$: Subcomplex, $r$: Integer}
    \State $\Omega \gets M_{\Call{Star}{C}}$ \Comment{Lattice map of the star of input subcomplex}
    \State $Directions \gets \Call{GetAllPrimalDirections}{r}$
    \State $V: \text{vector of boolean} \gets [0, \ldots, 0]$ \Comment{length $Size(Directions) \times \#C.pointels$}
    \State $low, high \gets \Call{BoundingBoxZ}{\Omega}$
    \ForAll{$\vv$ in $Directions$}
    \ForAll{shift $S$ in $\Omega$}
    \State $R \gets [low, high]$
    \ForAll{pair $P$ in $M_{\Call{Star}{[\mathbf{0},\mathbf{v}]}}$}
    \State $R \gets R \cap \Call{Translations}{P.intervals,\Omega[S + P.shift]}$
    \EndFor
    \State \Call{UpdateVisibility}{$V$, $R$}
    \EndFor
    \EndFor
    \State \Return $V$
    \EndFunction
  \end{algorithmic}
\end{algorithm}

\begin{figure}
  \centering
  \input{algorithm.tikz}
  \caption{Evolution of the visibility check algorithm for a $(2,1)$
    vector. Green is the vector lattice map, black is the figure
    lattice map, red are the current intervals of positions where
    the visibility is still possible. The last red intervals are
    the visible positions. We travel the lattice maps from
    bottom-up and the found visibilities are drawn on the
    uppest figure.}
  \label{fig:visibility-algorithm-evolution}
\end{figure}
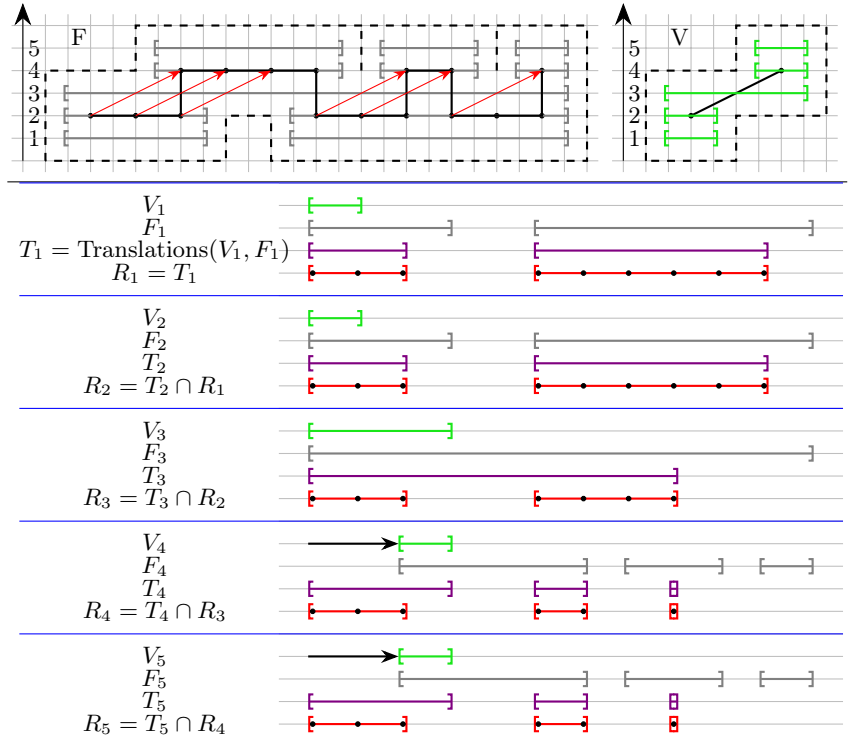

\begin{figure}
  \centering
  \begin{tabular}{c c}
    \includegraphics[width=0.4\textwidth]{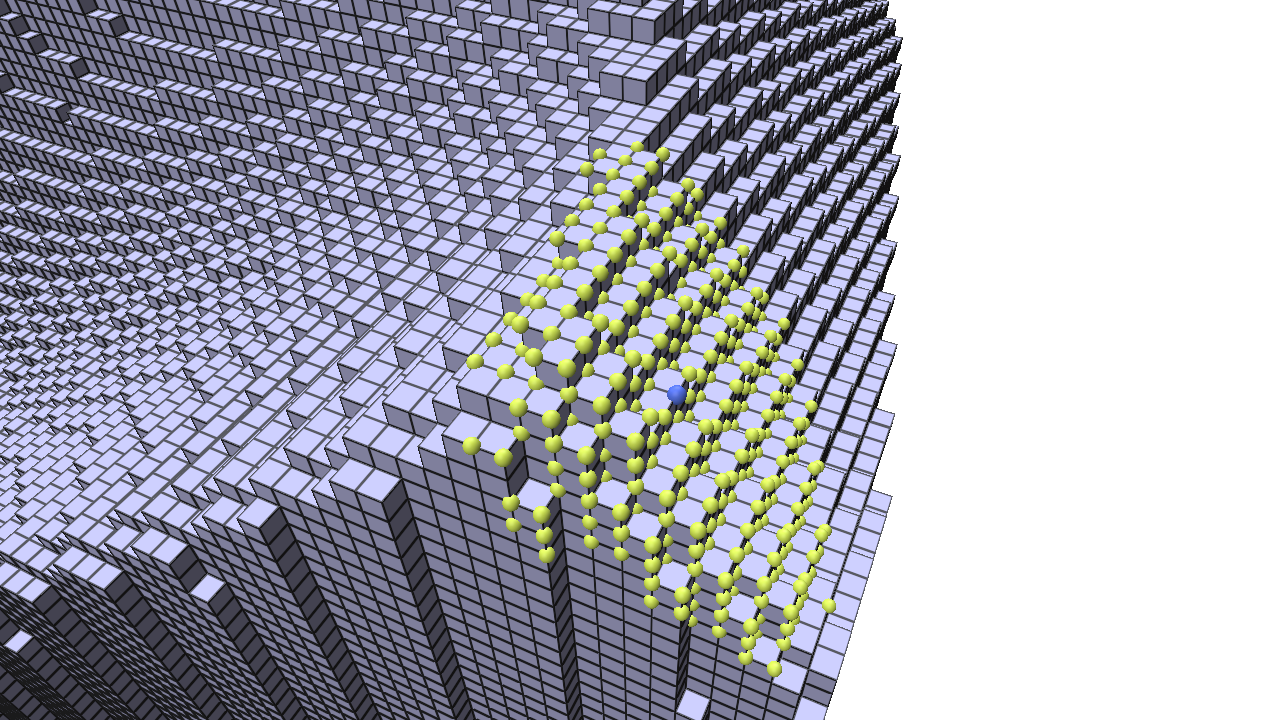} &
    \includegraphics[width=0.4\textwidth]{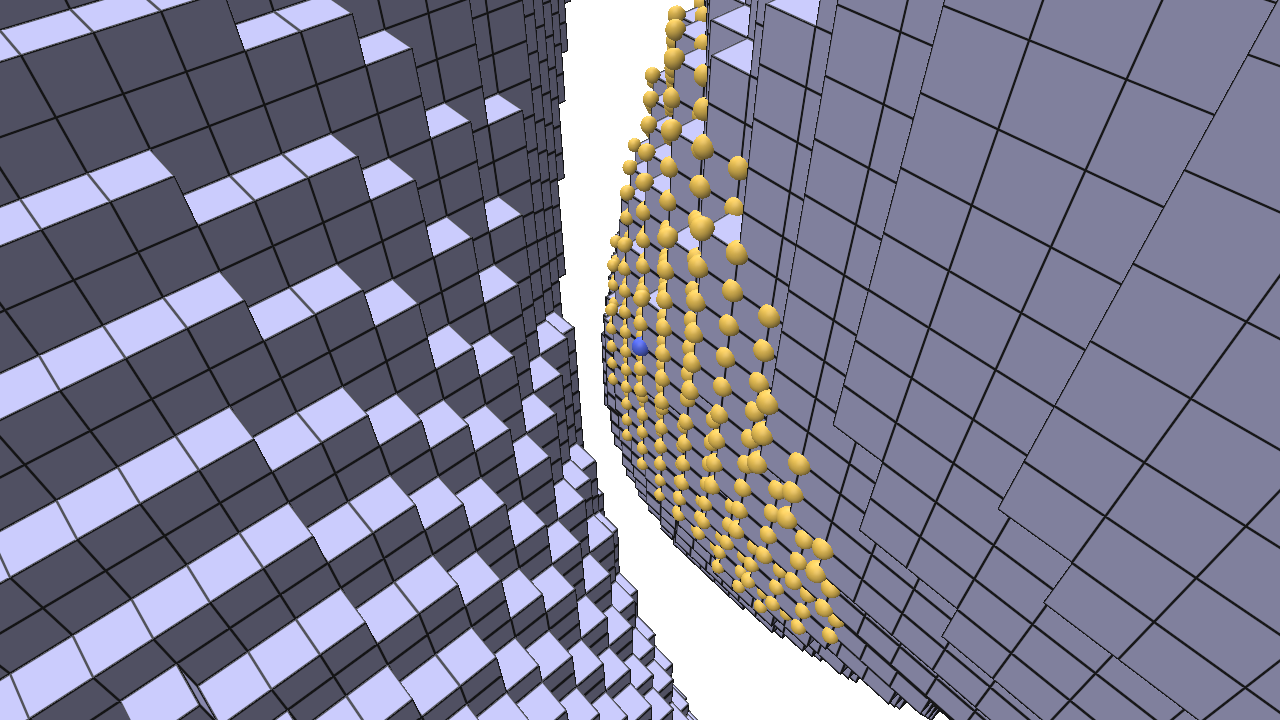}
  \end{tabular}
  \caption{Examples of visibilities from a source point:
    (left) the visibility is stopped at the sharp edge, (right) the
    visibility does not cross the gap between the two parts.}
  \label{fig:visibility-results}
\end{figure}

Figure~\ref{fig:visibility-algorithm-evolution} shows an example of
execution of this algorithm in an elementary case. It reduces to
determining all the possible inclusions by translation of sequences of
intervals in another sequence of intervals (note that for
pairwise visibility, it is enough to consider all the possible
inclusions by translation of one interval in another sequence of
intervals), and then intersecting progressively the results. Of
course, the process is stopped as soon as the intersection is
empty. Some results of visible points are displayed on
figure~\ref{fig:visibility-results} on 3d shapes.

\begin{algorithm}
  \caption{Given 2 lists of integer intervals $K$ and $L$, returns $K \cap L$}
  \label{alg:intersection}
  \begin{algorithmic}
    \Function{Intersection ($\cap$)}{\text{K, L}: Intervals}
    \State $R \gets \emptyset$; $k \gets 0$; $l \gets 0$;
    \While{$k < K.nbIntervals \And l < L.nbIntervals$}
    \State $[a,b] \gets K[k]$; $[c,d] \gets L[l]$
    \State $e \gets \max(a, c)$; $f \gets \min(b, d)$;
    \If{$e \leq f$}
    $R.append([e, f])$;
    \EndIf
    \If{$b \leq d$}
    $k \gets k+1$;
    \EndIf
    \If{$d \leq b$}
    $l \gets l+1$;
    \EndIf
    \EndWhile
    \State \Return $R$
    \EndFunction
  \end{algorithmic}
\end{algorithm}

To efficiently compute the intersection of two sequences of intervals
$K$ and $L$ (Algorithm~\ref{alg:intersection}), we go through the
first $2$ unvisited elements of the $2$ lists. If they do not overlap
(i.e.\ the start and end of the 1st interval are both smaller than the
start of the other), then we can discard the smaller interval. Else,
we construct the intersection of these 2 intervals as one of the
resulting intervals. We then can skip the interval with the smallest
end. If both intervals have the same end, then both intervals can be
skipped. Doing this computation until one of the lists is empty
returns the intersection of both lists of intervals.

\begin{figure}
  \centering
  \input{mean-distance.tikz}
  \caption{\label{fig:meanvisibility-gridstep} Mean distance of visibility as a function of the gridstep $h$ (see text).}
\end{figure}
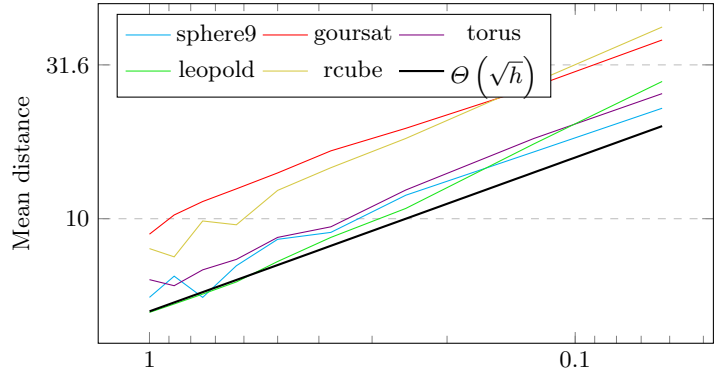

We have plotted on Figure~\ref{fig:meanvisibility-computationComplexity}
the running times to compute global visibility of both~\cite[Algorithm
  3]{lachaud:2022-jmiv} and our algorithm. Timings are very
similar. Our algorithm is faster for smaller maximal radii (up to 20),
while the other is slightly faster for bigger radii. We have also
measured the (discrete) average visibility distance along digitization
of smooth shapes (Figure~\ref{fig:meanvisibility-gridstep}). It follows
a law proportionnal to $\sqrt{h}$. As shown in the next section, for
real 3d images, a maximal radius below 10 is sufficient for normal and
curvature estimation.

\begin{figure}[t]
  \centering
  \input{pointelscomptime.tikz}
  \caption{ \label{fig:meanvisibility-computationComplexity}Computation
    time of visibility as a function of the number of pointels. We
    compare the running time of the naive breadth first visibility
    algorithm against our algorithm using intervals. We input the same
    maximal radius to both versions (10, 20, 30) and test it on 6
    different figures (goursat, torus, rcube, sphere9, leopold,
    ``D20'') with various gridsteps, ($\#$ pointels ranging from 520
    to 390235).}
\end{figure}
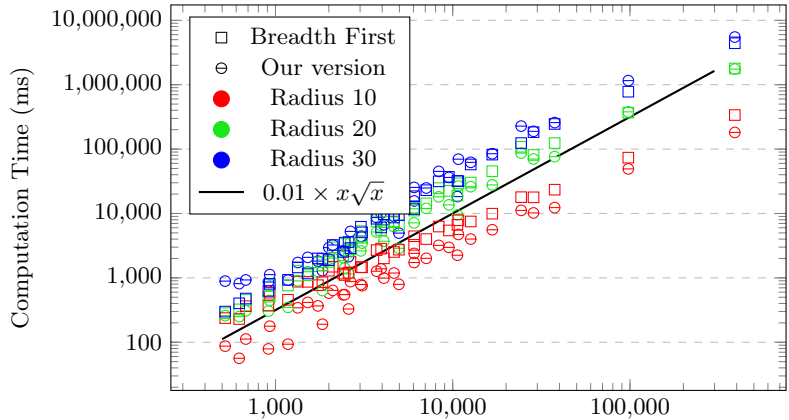

%% file: lattice.tikz
\begin{tabular}{c c}
  \begin{tikzpicture}[scale=0.75]
    \draw[step=0.5,lightgray,thin,xshift=-1cm,yshift=-1cm] (-0.25,1.75) grid (7.25,7.25);
    \filldraw[black] (1,4) circle (0.05);
    \filldraw[black] (1,5) circle (0.05);
    \filldraw[black] (2,5) circle (0.05);
    \filldraw[black] (3,5) circle (0.05);
    \filldraw[black] (3,4) circle (0.05);
    \filldraw[black] (3,3) circle (0.05);
    \filldraw[black] (3,2) circle (0.05);
    \filldraw[black] (2,2) circle (0.05);
    \filldraw[black] (1,2) circle (0.05);
    \filldraw[black] (4,2) circle (0.05);
    \draw[-{Stealth[length=3mm]}=black] (-1,1) -- (-1,6.5);
    \draw decorate [decoration={crosses,transform={rotate=45},shape size=1.5mm,segment length=21.5pt}] {(-1,1) -- (-1,7)};
    \draw[black] (1,4) -- (1,5) -- (3,5) -- (3,2) -- (4,2) -- (1,2) ;
    \draw[thick,dashed]  (0,3) -- (0,1) -- (5,1) -- (5,3) -- (4,3) -- (4,6) -- (0,6) -- (0,3) -- (2,3) -- (2,4);
  \end{tikzpicture} &
  \begin{tikzpicture}[scale=0.75]
    \foreach \y in {2,2.5,...,7} {
      \draw[thin, lightgray,xshift=-1cm,yshift=-1cm] (-0.25,\y) -- (7.25,\y);
    }
    \draw[-{Stealth[length=3mm]}=black] (-1,1) -- (-1,6.5);
    \draw decorate [decoration={crosses,transform={rotate=45},shape size=1.5mm,segment length=21.5pt}] {(-1,1) -- (-1,6.5)};
    \draw[thick,dashed]  (0,3) -- (0,1) -- (5,1) -- (5,3) -- (4,3) -- (4,6) -- (0,6) -- (0,3) -- (2,3) -- (2,4);
    \draw[thick, arrows = {Bracket[sharp]-Bracket[sharp]}] (0.4,5.5)--(3.6,5.5);
    \draw[thick, arrows = {Bracket[sharp]-Bracket[sharp]}] (0.4,5)--(3.6,5);
    \draw[thick, arrows = {Bracket[sharp]-Bracket[sharp]}] (0.4,4.5)--(3.6,4.5);
    \draw[thick, arrows = {Bracket[sharp]-Bracket[sharp]}] (0.4,4)--(1.6,4);
    \draw[thick, arrows = {Bracket[sharp]-Bracket[sharp]}] (2.4,4)--(3.6,4);
    \draw[thick, arrows = {Bracket[sharp]-Bracket[sharp]}] (0.4,3.5)--(1.6,3.5);
    \draw[thick, arrows = {Bracket[sharp]-Bracket[sharp]}] (2.4,3.5)--(3.6,3.5);
    \draw[thick, arrows = {Bracket[sharp]-Bracket[sharp]}] (2.4,3)--(3.6,3);
    \draw[thick, arrows = {Bracket[sharp]-Bracket[sharp]}] (0.4,2.5)--(4.6,2.5);
    \draw[thick, arrows = {Bracket[sharp]-Bracket[sharp]}] (0.4,2)--(4.6,2);
    \draw[thick, arrows = {Bracket[sharp]-Bracket[sharp]}] (0.4,1.5)--(4.6,1.5);
  \end{tikzpicture}\\[-8mm]
\end{tabular}

%% file: algorithm.tikz
\begin{tabular}{c c}
  \begin{tikzpicture}[scale=0.6]
    \foreach[count=\i] \y in {-0.5,0,...,1.5} {
      \draw (0,0) node at (-1.25,\y) {\i};
    }
    \draw[step=0.5,lightgray,thin,xshift=-1cm,yshift=-1cm] (-0.75,-0.25) grid (12.25,3.25);
    \draw[-{Stealth[length=3mm]}=black] (-1.5,-1) -- (-1.5,2.5);
    \filldraw[black] (0,0) circle (0.05);
    \filldraw[black] (1,0) circle (0.05);
    \filldraw[black] (2,0) circle (0.05);
    \filldraw[black] (2,1) circle (0.05);
    \filldraw[black] (3,1) circle (0.05);
    \filldraw[black] (4,1) circle (0.05);
    \filldraw[black] (5,1) circle (0.05);
    \filldraw[black] (5,0) circle (0.05);
    \filldraw[black] (6,0) circle (0.05);
    \filldraw[black] (7,0) circle (0.05);
    \filldraw[black] (7,1) circle (0.05);
    \filldraw[black] (8,1) circle (0.05);
    \filldraw[black] (8,0) circle (0.05);
    \filldraw[black] (9,0) circle (0.05);
    \filldraw[black] (10,0) circle (0.05);
    \filldraw[black] (10,1) circle (0.05);
    \draw[thick,gray, arrows = {Bracket[sharp]-Bracket[sharp]}] (-0.6,-0.5)--(2.6,-0.5);
    \draw[thick,gray, arrows = {Bracket[sharp]-Bracket[sharp]}] (4.4,-0.5)--(10.6,-0.5);
    \draw[thick,gray, arrows = {Bracket[sharp]-Bracket[sharp]}] (-0.6,0)--(2.6,0);
    \draw[thick,gray, arrows = {Bracket[sharp]-Bracket[sharp]}] (4.4,0)--(10.6,0);
    \draw[thick,gray, arrows = {Bracket[sharp]-Bracket[sharp]}] (-0.6,0.5)--(10.6,0.5);
    \draw[thick,gray, arrows = {Bracket[sharp]-Bracket[sharp]}] (1.4,1)--(5.6,1);
    \draw[thick,gray, arrows = {Bracket[sharp]-Bracket[sharp]}] (6.4,1)--(8.6,1);
    \draw[thick,gray, arrows = {Bracket[sharp]-Bracket[sharp]}] (9.4,1)--(10.6,1);
    \draw[thick,gray, arrows = {Bracket[sharp]-Bracket[sharp]}] (1.4,1.5)--(5.6,1.5);
    \draw[thick,gray, arrows = {Bracket[sharp]-Bracket[sharp]}] (6.4,1.5)--(8.6,1.5);
    \draw[thick,gray, arrows = {Bracket[sharp]-Bracket[sharp]}] (9.4,1.5)--(10.6,1.5);
    \draw[thick, black] (0,0) -- (2,0) -- (2,1) -- (5,1) -- (5,0) -- (7,0) -- (7,1) -- (8,1) -- (8,0) -- (10,0) -- (10,1);
    \draw[thick,dashed] (6,1) -- (6,2);
    \draw[thick,dashed] (9,1) -- (9,2);
    \draw[thick,dashed] (-1,-1) -- (-1,1) -- (1,1) -- (1,2) -- (11,2) -- (11,-1) -- (4,-1) -- (4,0) -- (3,0) -- (3,-1) -- (-1,-1);

    \draw (0,0) node at (-0.25,1.75) {F};

    \draw[red,arrows = {-Stealth[]}] (0,0)--(2,1);
    \draw[red,arrows = {-Stealth[]}] (1,0)--(3,1);
    \draw[red,arrows = {-Stealth[]}] (2,0)--(4,1);
    \draw[red,arrows = {-Stealth[]}] (5,0)--(7,1);
    \draw[red,arrows = {-Stealth[]}] (6,0)--(8,1);
    \draw[red,arrows = {-Stealth[]}] (8,0)--(10,1);

  \end{tikzpicture} &
  \begin{tikzpicture}[scale=0.6]
    \draw[step=0.5,lightgray,thin,xshift=-1cm,yshift=-1cm] (-0.75,-0.25) grid (4.25,3.25);
    \draw[-{Stealth[length=3mm]}=black] (-1.5,-1) -- (-1.5,2.5);
    \foreach[count=\i] \y in {-0.5,0,...,1.5} {
      \draw (0,0) node at (-1.25,\y) {\i};
    }
    \filldraw[black] (0,0) circle (0.05);
    \filldraw[black] (2,1) circle (0.05);
    \draw[thick, black] (0,0) -- (2,1);
    \draw[thick,dashed] (-1,-1) -- (-1,1) -- (1,1) -- (1,2) -- (3,2) -- (3,0) -- (1,0) --  (1,-1) -- (-1,-1);
    \draw[thick,MyGreen, arrows = {Bracket[sharp]-Bracket[sharp]}] (-0.6,-0.5)--(0.6,-0.5);
    \draw[thick,MyGreen, arrows = {Bracket[sharp]-Bracket[sharp]}] (-0.6,0)--(0.6,0);
    \draw[thick,MyGreen, arrows = {Bracket[sharp]-Bracket[sharp]}] (-0.6,0.5)--(2.6,0.5);
    \draw[thick,MyGreen, arrows = {Bracket[sharp]-Bracket[sharp]}] (1.4,1)--(2.6,1);
    \draw[thick,MyGreen, arrows = {Bracket[sharp]-Bracket[sharp]}] (1.4,1.5)--(2.6,1.5);

    \draw (0,0) node at (-0.25,1.75) {V};
  \end{tikzpicture} \\
  \hline
  \multicolumn{2}{c}{
    \begin{tikzpicture}[scale=0.6]
      \foreach \y in {0,0.5,...,12.5} {
        \draw[thin, lightgray] (-1.25,\y) -- (11.25,\y);
      }
      \foreach \y in {0,2.5,...,12.5} {
        \draw[blue] (-7,\y) -- (11.25,\y);
      }

      \foreach[count=\i] \y in {12,9.5,...,2} {
        \draw (0,0) node at (-4,\y) {$V_{\i}$};
      }
      \foreach[count=\i] \y in {11.5,9,...,1.5} {
        \draw (0,0) node at (-4,\y) {$F_{\i}$};
      }
      \draw (0,0) node at (-4,11) {$T_{1} = \text{Translations}(V_{1},F_{1})$};
      \foreach[count=\i] \y in {8.5,6,...,1} {
        \draw (0,0) node at (-4,\y) {$T_{\the\numexpr\i+1\relax}$};
      }
      \draw (0,0) node at (-4,10.5) {$R_1 = T_1$};

      \foreach[count=\i] \y in {8,5.5,...,0.5} {

        \draw (0,0) node at (-4,\y) {$R_{\the\numexpr\i+1\relax} = T_{\the\numexpr\i+1\relax} \cap R_{\i}$};
      }

      \draw[thick,MyGreen,arrows = {Bracket[sharp]-Bracket[sharp]}] (-0.6,12)--(0.6,12);
      \draw[thick,gray,arrows = {Bracket[sharp]-Bracket[sharp]}] (-0.6,11.5)--(2.6,11.5);
      \draw[thick,gray,arrows = {Bracket[sharp]-Bracket[sharp]}] (4.4,11.5)--(10.6,11.5);
      \draw[thick,violet,arrows = {Bracket[sharp]-Bracket[sharp]}] (-0.6,11)--(1.6,11);
      \draw[thick,violet,arrows = {Bracket[sharp]-Bracket[sharp]}] (4.4,11)--(9.6,11);
      \draw[thick,red,arrows = {Bracket[sharp]-Bracket[sharp]}] (-0.6,10.5)--(1.6,10.5);
      \draw[thick,red,arrows = {Bracket[sharp]-Bracket[sharp]}] (4.4,10.5)--(9.6,10.5);
      \filldraw[black] (-0.5,10.5) circle (0.05);
      \filldraw[black] (0.5,10.5) circle (0.05);
      \filldraw[black] (1.5,10.5) circle (0.05);
      \filldraw[black] (4.5,10.5) circle (0.05);
      \filldraw[black] (5.5,10.5) circle (0.05);
      \filldraw[black] (6.5,10.5) circle (0.05);
      \filldraw[black] (7.5,10.5) circle (0.05);
      \filldraw[black] (8.5,10.5) circle (0.05);
      \filldraw[black] (9.5,10.5) circle (0.05);

      \draw[thick,MyGreen,arrows = {Bracket[sharp]-Bracket[sharp]}] (-0.6,9.5)--(0.6,9.5);
      \draw[thick,gray,arrows = {Bracket[sharp]-Bracket[sharp]}] (-0.6,9)--(2.6,9);
      \draw[thick,gray,arrows = {Bracket[sharp]-Bracket[sharp]}] (4.4,9)--(10.6,9);
      \draw[thick,violet,arrows = {Bracket[sharp]-Bracket[sharp]}] (-0.6,8.5)--(1.6,8.5);
      \draw[thick,violet,arrows = {Bracket[sharp]-Bracket[sharp]}] (4.4,8.5)--(9.6,8.5);
      \draw[thick,red,arrows = {Bracket[sharp]-Bracket[sharp]}] (-0.6,8)--(1.6,8);
      \draw[thick,red,arrows = {Bracket[sharp]-Bracket[sharp]}] (4.4,8)--(9.6,8);
      \filldraw[black] (-0.5,8) circle (0.05);
      \filldraw[black] (0.5,8) circle (0.05);
      \filldraw[black] (1.5,8) circle (0.05);
      \filldraw[black] (4.5,8) circle (0.05);
      \filldraw[black] (5.5,8) circle (0.05);
      \filldraw[black] (6.5,8) circle (0.05);
      \filldraw[black] (7.5,8) circle (0.05);
      \filldraw[black] (8.5,8) circle (0.05);
      \filldraw[black] (9.5,8) circle (0.05);

      \draw[thick,MyGreen,arrows = {Bracket[sharp]-Bracket[sharp]}] (-0.6,7)--(2.6,7);
      \draw[thick,gray,arrows = {Bracket[sharp]-Bracket[sharp]}] (-0.6,6.5)--(10.6,6.5);
      \draw[thick,violet,arrows = {Bracket[sharp]-Bracket[sharp]}] (-0.6,6)--(7.6,6);
      \draw[thick,red,arrows = {Bracket[sharp]-Bracket[sharp]}] (-0.6,5.5)--(1.6,5.5);
      \draw[thick,red,arrows = {Bracket[sharp]-Bracket[sharp]}] (4.4,5.5)--(7.6,5.5);
      \filldraw[black] (-0.5,5.5) circle (0.05);
      \filldraw[black] (0.5,5.5) circle (0.05);
      \filldraw[black] (1.5,5.5) circle (0.05);
      \filldraw[black] (4.5,5.5) circle (0.05);
      \filldraw[black] (5.5,5.5) circle (0.05);
      \filldraw[black] (6.5,5.5) circle (0.05);
      \filldraw[black] (7.5,5.5) circle (0.05);

      \draw[thick,black,arrows = {-Stealth[]}] (-0.6,4.5)--(1.4,4.5);
      \draw[thick,MyGreen,arrows = {Bracket[sharp]-Bracket[sharp]}] (1.4,4.5)--(2.6,4.5);
      \draw[thick,gray,arrows = {Bracket[sharp]-Bracket[sharp]}] (1.4,4)--(5.6,4);
      \draw[thick,gray,arrows = {Bracket[sharp]-Bracket[sharp]}] (6.4,4)--(8.6,4);
      \draw[thick,gray,arrows = {Bracket[sharp]-Bracket[sharp]}] (9.4,4)--(10.6,4);
      \draw[thick,violet,arrows = {Bracket[sharp]-Bracket[sharp]}] (-0.6,3.5)--(2.6,3.5);
      \draw[thick,violet,arrows = {Bracket[sharp]-Bracket[sharp]}] (4.4,3.5)--(5.6,3.5);
      \draw[thick,violet,arrows = {Bracket[sharp]-Bracket[sharp]}] (7.4,3.5)--(7.6,3.5);
      \draw[thick,red,arrows = {Bracket[sharp]-Bracket[sharp]}] (-0.6,3)--(1.6,3);
      \draw[thick,red,arrows = {Bracket[sharp]-Bracket[sharp]}] (4.4,3)--(5.6,3);
      \draw[thick,red,arrows = {Bracket[sharp]-Bracket[sharp]}] (7.4,3)--(7.6,3);
      \filldraw[black] (-0.5,3) circle (0.05);
      \filldraw[black] (0.5,3) circle (0.05);
      \filldraw[black] (1.5,3) circle (0.05);
      \filldraw[black] (4.5,3) circle (0.05);
      \filldraw[black] (5.5,3) circle (0.05);
      \filldraw[black] (7.5,3) circle (0.05);

      \draw[thick,black,arrows = {-Stealth[]}] (-0.6,2)--(1.4,2);
      \draw[thick,MyGreen,arrows = {Bracket[sharp]-Bracket[sharp]}] (1.4,2)--(2.6,2);
      \draw[thick,gray,arrows = {Bracket[sharp]-Bracket[sharp]}] (1.4,1.5)--(5.6,1.5);
      \draw[thick,gray,arrows = {Bracket[sharp]-Bracket[sharp]}] (6.4,1.5)--(8.6,1.5);
      \draw[thick,gray,arrows = {Bracket[sharp]-Bracket[sharp]}] (9.4,1.5)--(10.6,1.5);
      \draw[thick,violet,arrows = {Bracket[sharp]-Bracket[sharp]}] (-0.6,1)--(2.6,1);
      \draw[thick,violet,arrows = {Bracket[sharp]-Bracket[sharp]}] (4.4,1)--(5.6,1);
      \draw[thick,violet,arrows = {Bracket[sharp]-Bracket[sharp]}] (7.4,1)--(7.6,1);
      \draw[thick,red,arrows = {Bracket[sharp]-Bracket[sharp]}] (-0.6,0.5)--(1.6,0.5);
      \draw[thick,red,arrows = {Bracket[sharp]-Bracket[sharp]}] (4.4,0.5)--(5.6,0.5);
      \draw[thick,red,arrows = {Bracket[sharp]-Bracket[sharp]}] (7.4,0.5)--(7.6,0.5);
      \filldraw[black] (-0.5,0.5) circle (0.05);
      \filldraw[black] (0.5,0.5) circle (0.05);
      \filldraw[black] (1.5,0.5) circle (0.05);
      \filldraw[black] (4.5,0.5) circle (0.05);
      \filldraw[black] (5.5,0.5) circle (0.05);
      \filldraw[black] (7.5,0.5) circle (0.05);

  \end{tikzpicture}}
\end{tabular}

%% file: pointelscomptime.tikz
\begin{tikzpicture}
    \centering
    \begin{axis}[
        width=0.8\textwidth,
        height=0.55\textwidth,
        ylabel={Computation Time (ms)},
        legend pos=north west,
        ymajorgrids=true,
        grid style=dashed,
        ymode=log,
        xmode=log,
        log ticks with fixed point,
    ]

        \addlegendimage{
            black,
            only marks,
            mark=square,
            mark options={solid}
        }
        \addlegendentry{Breadth First}
        \addlegendimage{
            black,
            only marks,
            mark=halfcircle,
            mark options={solid}
        }
        \addlegendentry{Our version}

        \addlegendimage{only marks, mark=*, color=red, mark size=3pt}
        \addlegendentry{Radius 10}
        \addlegendimage{only marks, mark=*, color=MyGreen, mark size=3pt}
        \addlegendentry{Radius 20}
        \addlegendimage{only marks, mark=*, color=blue, mark size=3pt}
        \addlegendentry{Radius 30}

        \addlegendimage{color=black,thick};
        \addlegendentry{$0.01\times x\sqrt{x}$}

        \addplot+[
            red,
            only marks,
            mark=halfcircle,
            mark options={solid}
        ] table[row sep=crcr] {
            10624 2246.69\\
            4968 793.426\\
            2584 328.535\\
            1840 190.401\\
            1176 93.2502\\
            912 78.7981\\
            624 56.3769\\
            28568 10193.5\\
            12632 4004.92\\
            7088 2010.63\\
            4664 1188.73\\
            3080 755.754\\
            2408 545.347\\
            1736 366.541\\
            24320 11169.1\\
            10760 4720.44\\
            6056 2397.79\\
            3944 1380.38\\
            2648 874.979\\
            2000 580.0\\
            1520 415.373\\
            8336 3196.55\\
            3720 1268.46\\
            2104 639.305\\
            1336 342.873\\
            928 177.54\\
            680 112.142\\
            520 87.0747\\
            37592 12335.6\\
            16712 5582.38\\
            9512 2976.21\\
            6056 1731.32\\
            4088 990.78\\
            3032 788.507\\
            2456 548.66\\
            97844 49226.3\\
            390235 181322\\
        };
        \addplot+[
            MyGreen,
            only marks,
            mark=halfcircle,
            mark options={solid}
        ] table[row sep=crcr] {
            10624 9522.43\\
            4968 2876.81\\
            2584 1152.25\\
            1840 635.332\\
            1176 347.973\\
            912 304.801\\
            624 250.126\\
            28568 70137.9\\
            12632 26063.0\\
            7088 11985.8\\
            4664 6239.51\\
            3080 3468.31\\
            2408 2059.84\\
            1736 1254.2\\
            24320 86616.9\\
            10760 26605.7\\
            6056 11813.5\\
            3944 6516.92\\
            2648 3564.03\\
            2000 1949.32\\
            1520 1244.89\\
            8336 18125.5\\
            3720 5157.11\\
            2104 2029.09\\
            1336 896.186\\
            928 446.136\\
            680 306.651\\
            520 259.078\\
            37592 76316.3\\
            16712 27747.3\\
            9512 13608.6\\
            6056 7140.03\\
            4088 3704.52\\
            3032 2624.2\\
            2456 1422.6\\
            97844 378906\\
            390235 1747780\\
        };
        \addplot+[
            blue,
            only marks,
            mark=halfcircle,
            mark options={solid}
        ] table[row sep=crcr] {
            10624 18283.5\\
            4968 4957.93\\
            2584 2133.36\\
            1840 1292.53\\
            1176 922.51\\
            912 863.731\\
            624 813.638\\
            28568 188784.0\\
            12632 62237.1\\
            7088 24874.9\\
            4664 9043.39\\
            3080 4407.09\\
            2408 2641.04\\
            1736 1820.31\\
            24320 227432.0\\
            10760 69706.8\\
            6056 25408.9\\
            3944 10094.7\\
            2648 5269.56\\
            2000 2948.16\\
            1520 2082.02\\
            8336 45045.4\\
            3720 8638.42\\
            2104 3238.48\\
            1336 1731.9\\
            928 1122.69\\
            680 928.084\\
            520 890.592\\
            37592 257981.0\\
            16712 84729.7\\
            9512 36298.8\\
            6056 15640.2\\
            4088 6662.17\\
            3032 4431.71\\
            2456 2621.98\\
            97844 1157400\\
            390235 5493200\\
        };
        \addplot+[
            red,
            only marks,
            mark=square,
            mark options={solid}
        ] table[row sep=crcr] {
            10624 6658.15\\
            4968 2740.74\\
            2584 1198.02\\
            1840 853.906\\
            1176 456.784\\
            912 373.028\\
            624 229.479\\
            28568 17837.4\\
            12632 7512.6\\
            7088 4005.5\\
            4664 2522.87\\
            3080 1458.22\\
            2408 1164.01\\
            1736 770.449\\
            24320 17922.8\\
            10760 7845.85\\
            6056 4431.68\\
            3944 2840.96\\
            2648 1774.39\\
            2000 1292.66\\
            1520 867.89\\
            8336 6268.17\\
            3720 2697.19\\
            2104 1489.4\\
            1336 871.987\\
            928 588.548\\
            680 363.434\\
            520 237.277\\
            37592 23416.9\\
            16712 9923.63\\
            9512 5446.57\\
            6056 3256.35\\
            4088 2008.14\\
            3032 1467.76\\
            2456 1107.06\\
            97844 73606.0\\
            390235 337851.0\\
        };
        \addplot+[
            MyGreen,
            only marks,
            mark=square,
            mark options={solid}
        ] table[row sep=crcr] {
            10624 18932.3\\
            4968 6555.12\\
            2584 2621.4\\
            1840 1571.18\\
            1176 759.752\\
            912 510.296\\
            624 323.584\\
            28568 81041.4\\
            12632 30169.2\\
            7088 14369.6\\
            4664 7318.29\\
            3080 4377.33\\
            2408 2255.89\\
            1736 1743.95\\
            24320 108696.0\\
            10760 29433.6\\
            6056 11092.9\\
            3944 5535.89\\
            2648 2759.14\\
            2000 1770.88\\
            1520 1103.44\\
            8336 24782.2\\
            3720 7521.02\\
            2104 2976.49\\
            1336 1296.38\\
            928 777.747\\
            680 455.317\\
            520 283.182\\
            37592 124066.0\\
            16712 45347.2\\
            9512 22445.1\\
            6056 12308.8\\
            4088 7370.63\\
            3032 5317.86\\
            2456 3356.28\\
            97844 368339.0\\
            390235 1784570.0\\
            };
        \addplot+[
            blue,
            only marks,
            mark=square,
            mark options={solid}
        ] table[row sep=crcr] {
            10624 32025.4\\
            4968 9426.13\\
            2584 3596.87\\
            1840 1975.03\\
            1176 930.202\\
            912 626.321\\
            624 402.24\\
            28568 183072.0\\
            12632 57696.6\\
            7088 22998.7\\
            4664 8654.47\\
            3080 5090.0\\
            2408 2538.87\\
            1736 2004.77\\
            24320 124004.0\\
            10760 31699.4\\
            6056 11704.3\\
            3944 6068.31\\
            2648 2885.91\\
            2000 1888.12\\
            1520 1170.45\\
            8336 31540.8\\
            3720 8280.03\\
            2104 3236.58\\
            1336 1451.9\\
            928 807.249\\
            680 477.556\\
            520 299.46\\
            37592 245584.0\\
            16712 82093.2\\
            9512 37091.0\\
            6056 18530.7\\
            4088 9711.66\\
            3032 6214.95\\
            2456 3523.63\\
            97844 778491.0\\
            390235 4409340.0\\
            };
        \addplot[
            color=black,
            thick,
            domain=500:300000,
            samples=200,
        ] {0.01*x^(1.5)};
    \end{axis}
\end{tikzpicture}

%% file: mean-distance.tikz
  \begin{tikzpicture}
    \centering
    \begin{axis}[
        width=0.8\textwidth,
        height=0.5\textwidth,
        legend columns=3,
        ylabel={Mean distance},
        x dir=reverse,
        legend pos=north west,
        ymajorgrids=true,
        grid style=dashed,
        ymax=50,
        ymode=log,
        xmode=log,
        log ticks with fixed point,
      ]

      \addplot[
        color=cyan,
      ] coordinates {
        (0.0625,3*7.61948)(0.125,3*5.49755)(0.25,3*3.97418)(0.375,3*3.00921)(0.5,3*2.85702)(0.625,3*2.34448)(0.75,3*1.85023)(0.875,3*2.16868)(1.0,3*1.85023)
      };
      \addlegendentry{sphere9}
      \addplot[
        color=red,
      ] coordinates {
        (0.0625,38.1045)(0.125,26.8765)(0.25,19.6747)(0.375,16.6192)(0.5,14.0867)(0.625,12.5062)(0.75,11.3734)(0.875,10.2845)(1,8.90781)
      };
      \addlegendentry{goursat}
      \addplot[
        color=violet,
      ] coordinates {
        (0.0625,25.5024)(0.125,18.2364)(0.25,12.3975)(0.375,9.41303)(0.5,8.69558)(0.625,7.37667)(0.75,6.82131)(0.875,6.05735)(1,6.33877)
      };
      \addlegendentry{torus}
      \addplot[
        color=MyGreen,
      ] coordinates {
        (0.0625,27.9287)(0.125,17.5104)(0.25,10.8007)(0.375,8.69514)(0.5,7.26074)(0.625,6.2309)(0.75,5.6897)(0.875,5.28197)(1,4.96098)
      };
      \addlegendentry{leopold}
      \addplot[
        color=yellow!80!black,
      ] coordinates {
        (0.0625,41.9816)(0.125,27.7104)(0.25,18.2339)(0.375,14.6475)(0.5,12.3618)(0.625,9.55647)(0.75,9.83208)(0.875,7.51861)(1,7.99732)
      };
      \addlegendentry{rcube}
      \addplot [
        color=black,
        thick,
        domain=0.0625:1,
        samples=100,
      ] {5*x^(-0.5)};
      \addlegendentry{$\Theta\left(\sqrt{h}\right)$} %
    \end{axis}
  \end{tikzpicture}

%% file: normalsandcurvature.tex
\newcommand{\Kernel}[1]{\ensuremath{w_{\sigma}(#1)}}

\paragraph{Saliency-aware normal estimator.}
We propose a new normal estimator on digital surfaces, which uses
the visibility of the point of interest while taking into account
a user-given scale $\sigma$. If $\sigma$ is proportionnal to the
average distance between visible points (so some
$\Theta\left(\sqrt{h}\right)$), then this estimator is observed to be
multigrid convergent along digitization of smooth shapes. However,
since the computation window is limited by the visibility, it
better approximates the geometry near sharp or salient features.

More precisely, we choose a Gaussian function
$\Kernel{x}\coloneqq e^{-\frac{x^2}{2\sigma^2}}$ as a weight kernel. Let
$V_p$ be the set of visible points from the point $p$. Let $c_p$
be the weighted centroid of the visible points around $p$,
i.e. $c_p \coloneqq \frac{\sum_{q \in V_p} \Kernel{\|q-p\|}q}{\sum_{q \in
V_p} \Kernel{\|q-p\|}}$. We form the weighted covariance matrix
$\mathcal{V}_p$ of the points $V_p$ as:
\begin{equation}
    \mathcal{V}_p = \sum_{q \in V_p} \Kernel{\|q-p\|}(q - c_p)(q - c_p)^T.
\end{equation}
The \emph{visibility normal} $\vec{n}(p)$ of point $p$ at scale $\sigma$ is defined
as the first eigenvector of the covariance matrix $\mathcal{V}_p$
of the visible points, corresponding to its smallest
eigenvalue. Its orientation is chosen to point in the same
direction as the average of the trivial normals to the surfels
touching the point $p$.

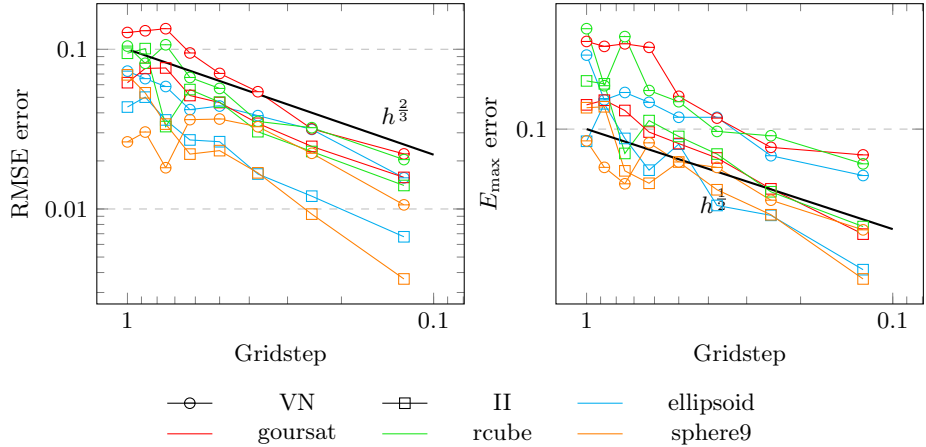
\begin{figure}
    \centering
    \begin{tabular}{c c}
        \begin{tikzpicture}
            \centering
            \begin{axis}[
                width=0.53\textwidth,
                xlabel={Gridstep},
                ylabel={RMSE error},
                x dir=reverse,
                legend pos=north east,
                ymajorgrids=true,
                grid style=dashed,
                ymode=log,
                xmode=log,
                log ticks with fixed point,
            ]
                \addplot[
                    color=cyan,
                    mark=halfcircle
                ] coordinates {
                    (0.125,0.0156736)(0.250,0.031503)(0.375,0.0384303)(0.500,0.0441086)(0.625,0.0418896)(0.750,0.0584604)(0.875,0.065503)(1.000,0.0729735)
                };

                \addplot[
                    color=cyan,
                    mark=square
                ] coordinates {
                    (0.125,0.00669781)(0.250,0.0120283)(0.375,0.0165559)(0.500,0.0263742)(0.625,0.0270529)(0.750,0.0361097)(0.875,0.0501562)(1.000,0.04348)
                };
                \addplot[
                    color=red,
                    mark=halfcircle
                ] coordinates {
                    (0.125,0.0221236)(0.250,0.0317159)(0.375,0.0542242)(0.500,0.0706101)(0.625,0.09481)(0.750,0.134544)(0.875,0.130578)(1.000,0.127371)
                };

                \addplot[
                    color=red,
                    mark=square
                ] coordinates {
                    (0.125,0.0157673)(0.250,0.024652)(0.375,0.0343723)(0.500,0.0463823)(0.625,0.051336)(0.750,0.0763857)(0.875,0.0758734)(1.000,0.0617788)

                };
                \addplot[
                    color=MyGreen,
                    mark=halfcircle
                ] coordinates {
                    (0.125,0.0203851)(0.250,0.032289)(0.375,0.0353311)(0.500,0.0568694)(0.625,0.0665957)(0.750,0.106711)(0.875,0.0817172)(1.000,0.104688)
                };

                \addplot[
                    color=MyGreen,
                    mark=square
                ] coordinates {
                    (0.125,0.0140342)(0.250,0.0229046)(0.375,0.0304815)(0.500,0.0466803)(0.625,0.0559396)(0.750,0.0328143)(0.875,0.100956)(1.000,0.0944342)

                };
                \addplot[
                    color=orange,
                    mark=halfcircle
                ] coordinates {
                    (0.125,0.0106125)(0.250,0.0223176)(0.375,0.032738)(0.500,0.0365776)(0.625,0.0362568)(0.750,0.0181232)(0.875,0.0303189)(1.000,0.0263066)
                };

                \addplot[
                    color=orange,
                    mark=square
                ] coordinates {
                    (0.125,0.00364692)(0.250,0.00928532)(0.375,0.0168347)(0.500,0.0231946)(0.625,0.0220619)(0.750,0.0340007)(0.875,0.0533849)(1.000,0.0691569)

                };
                \addplot[
                    color=black,
                    thick,
                    domain=0.1:1,
                    samples=200,
                ] {0.1*x^(0.66)} node[pos=0.2, anchor=south west] {$h^{\frac{2}{3}}$};
            \end{axis}
        \end{tikzpicture} &
        \begin{tikzpicture}
            \centering
            \begin{axis}
                [
                width=0.53\textwidth,
                xlabel={Gridstep},
                ylabel={$E_{\max}$ error},
                x dir=reverse,
                legend pos=north east,
                ymajorgrids=true,
                grid style=dashed,
                ymode=log,
                xmode=log,
                log ticks with fixed point,
                ]
                \addplot[
                    color=cyan,
                    mark=halfcircle
                ] coordinates {
                    (0.125,0.0586128)(0.250,0.0736204)(0.375,0.114564)(0.500,0.114592)(0.625,0.136021)(0.750,0.152418)(0.875,0.139939)(1.000,0.234379)
                };

                \addplot[
                    color=cyan,
                    mark=square
                ] coordinates {
                    (0.125,0.0198319)(0.250,0.0370639)(0.375,0.0415253)(0.500,0.0840702)(0.625,0.0623331)(0.750,0.0898241)(0.875,0.133761)(1.000,0.0869342)
                };
                \addplot[
                    color=red,
                    mark=halfcircle
                ] coordinates {
                    (0.125,0.0743185)(0.250,0.0810657)(0.375,0.113504)(0.500,0.145859)(0.625,0.256274)(0.750,0.266855)(0.875,0.259101)(1.000,0.274269)
                };

                \addplot[
                    color=red,
                    mark=square
                ] coordinates {

                    (0.125,0.02986)(0.250,0.0502663)(0.375,0.0717319)(0.500,0.0846663)(0.625,0.0966577)(0.750,0.123886)(0.875,0.139487)(1.000,0.132201)
                };
                \addplot[
                    color=MyGreen,
                    mark=halfcircle
                ] coordinates {
                    (0.125,0.0670488)(0.250,0.0926269)(0.375,0.0973621)(0.500,0.137059)(0.625,0.155913)(0.750,0.290065)(0.875,0.164799)(1.000,0.317625)
                };

                \addplot[
                    color=MyGreen,
                    mark=square
                ] coordinates {

                    (0.125,0.0325697)(0.250,0.0486132)(0.375,0.0750567)(0.500,0.0918979)(0.625,0.11031)(0.750,0.0755826)(0.875,0.168386)(1.000,0.174149)
                };
                \addplot[
                    color=orange,
                    mark=halfcircle
                ] coordinates {
                    (0.125,0.0313599)(0.250,0.044124)(0.375,0.0643389)(0.500,0.0684951)(0.625,0.0853552)(0.750,0.0532733)(0.875,0.064386)(1.000,0.0873334)
                };

                \addplot[
                    color=orange,
                    mark=square
                ] coordinates {

                    (0.125,0.0178225)(0.250,0.0373191)(0.375,0.0497523)(0.500,0.0684615)(0.625,0.0536034)(0.750,0.0618511)(0.875,0.129302)(1.000,0.12792)
                };
                \addplot[
                    color=black,
                    thick,
                    domain=0.1:1,
                    samples=200,
                ] {0.1*x^(0.5)} node[pos=0.5, anchor=north east] {$h^{\frac{1}{2}}$};
            \end{axis}
        \end{tikzpicture} \\
        \multicolumn{2}{c}{
            \begin{tikzpicture}
                \begin{axis}[
                    hide axis,
                    xmin=0, xmax=1,
                    ymin=0, ymax=1,
                    legend columns=3,
                    legend style={
                        at={(0.5,-0.2)},
                        anchor=north,
                        draw=none,
                        column sep=15pt,
                    }
                ]
                    \addlegendimage{mark=halfcircle}
                    \addlegendentry{VN}
                    \addlegendimage{mark=square}
                    \addlegendentry{II}
                    \addlegendimage{color=cyan}
                    \addlegendentry{ellipsoid}
                    \addlegendimage{color=red}
                    \addlegendentry{goursat}
                    \addlegendimage{color=MyGreen}
                    \addlegendentry{rcube}
                    \addlegendimage{color=orange}
                    \addlegendentry{sphere9}

                \end{axis}
            \end{tikzpicture}
        } \\
    \end{tabular}
    \caption{Error RMSE and $E_{\max}$ as a function of grid resolution}
    \label{fig:errors-normals}
\end{figure}

\begin{figure}
    \centering
    \begin{tabular}{|c||c|c|}
        \hline
        Normals & With cube edges & Flat (no shading) \\
        \hline
        \hline
        \raisebox{18mm}{II} &
        \includegraphics[width=0.3\textwidth]{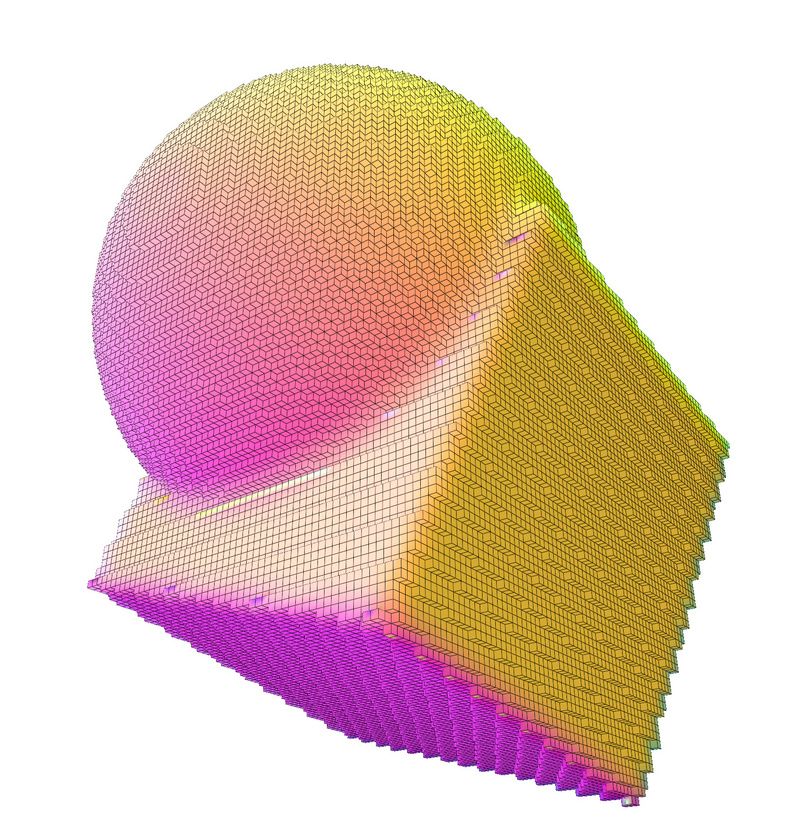} &
        \includegraphics[width=0.3\textwidth]{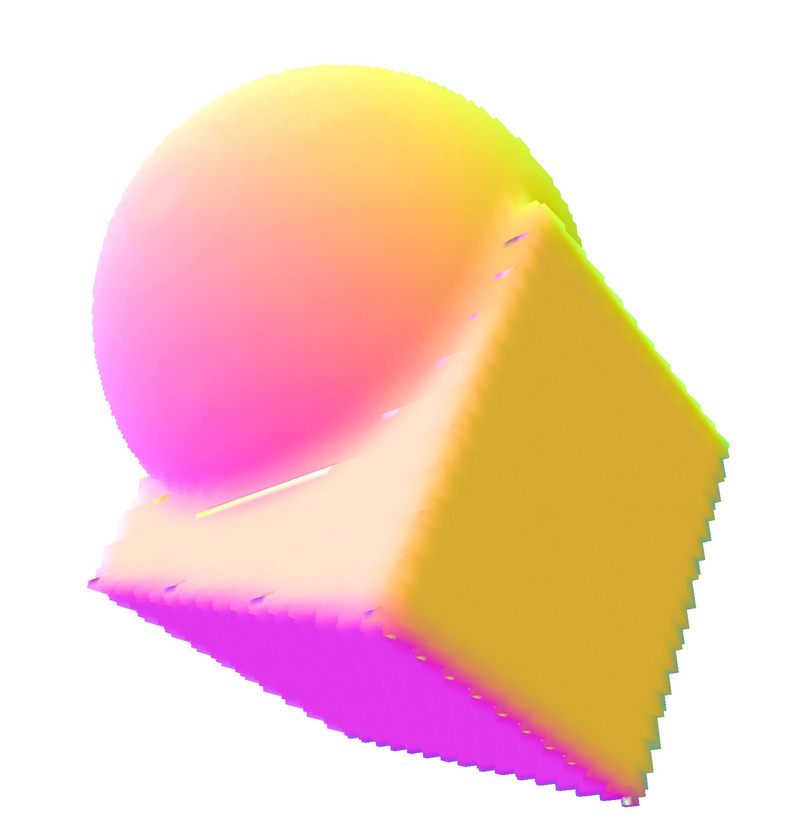} \\
        \hline
        \raisebox{18mm}{Ours} &
        \includegraphics[width=0.3\textwidth]{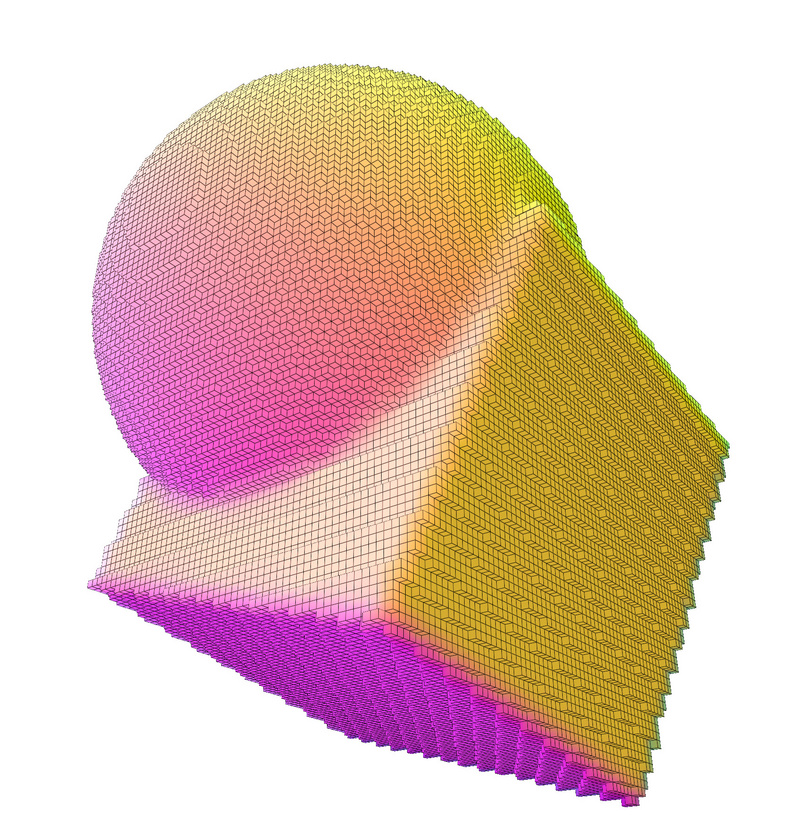} &
        \includegraphics[width=0.3\textwidth]{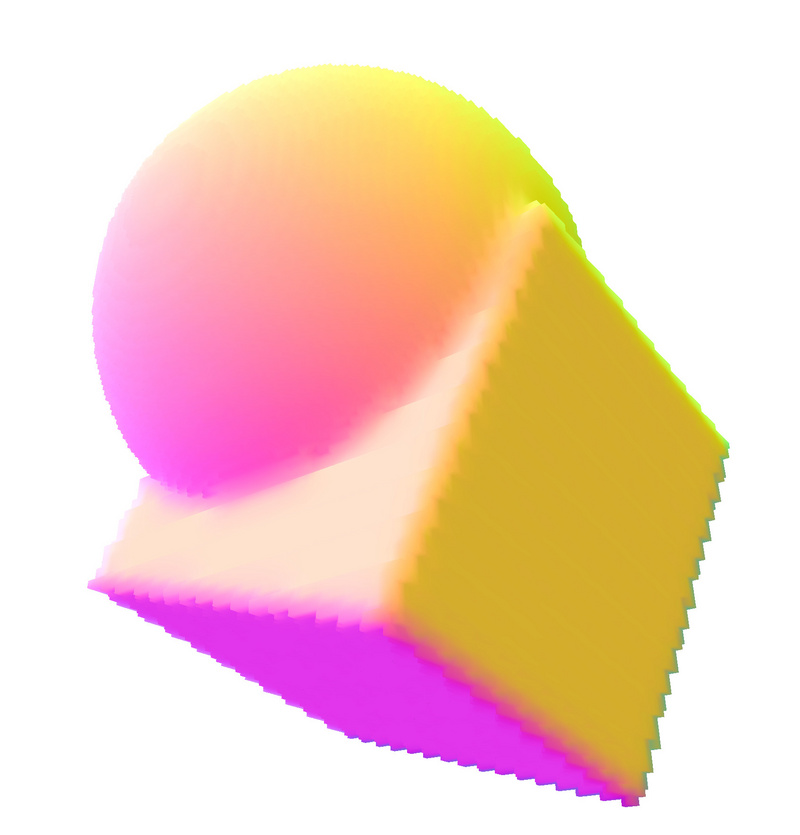} \\
        \hline
        \raisebox{18mm}{II} &
        \includegraphics[width=0.43\textwidth]{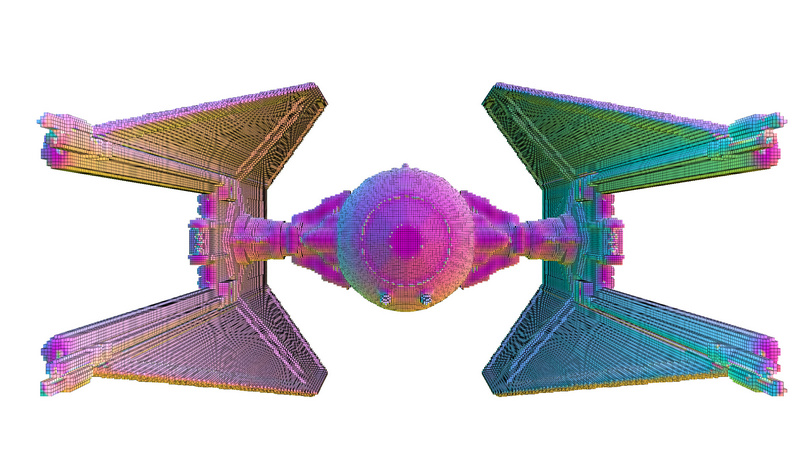} &
        \includegraphics[width=0.43\textwidth]{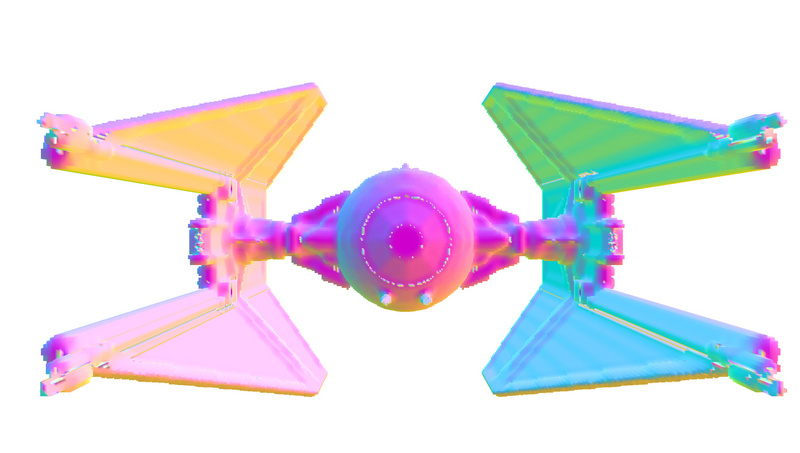} \\
        \hline
        \raisebox{18mm}{Ours} &
        \includegraphics[width=0.43\textwidth]{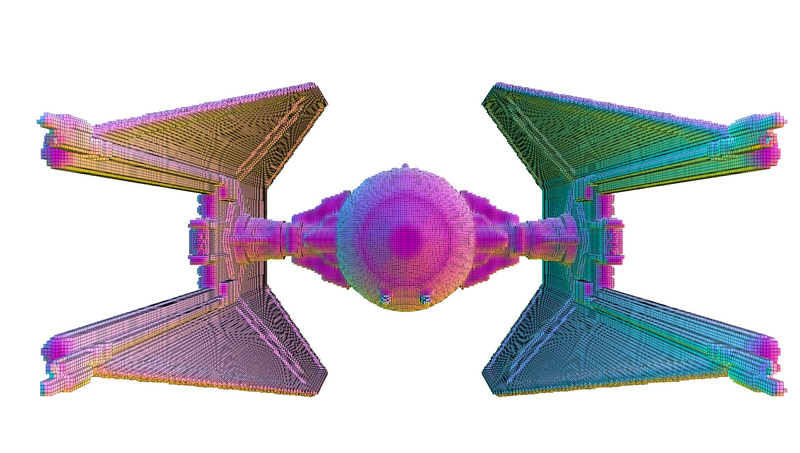} &
        \includegraphics[width=0.43\textwidth]{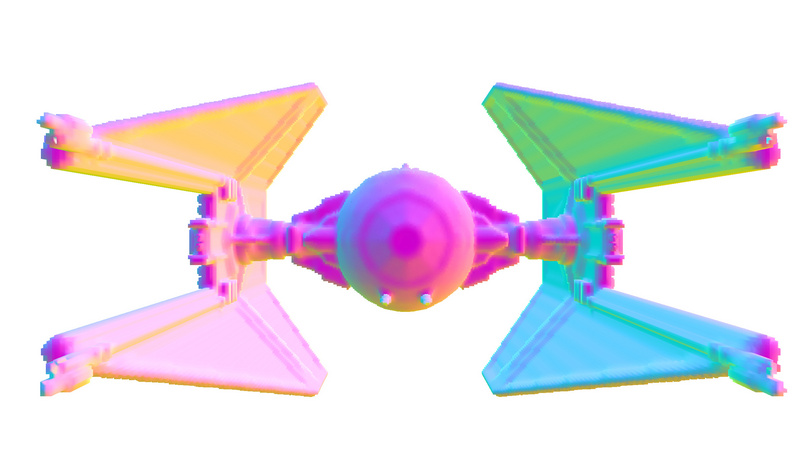} \\
        \hline
    \end{tabular}
    \caption{Examples of normals computed on a cps, then on a
    tie as a color map. First set displays II normals, second set
    uses the visibility algorithm to compute them. The normals are colored
    according to their orientation as $0.5 \cdot (n_i + \mathbbm{1})$. The
    right images are the smooth version of the left images. ``Integral
    Invariant normals'' are computed using a radius of $r=4.5$, while
    our normal estimator are computed using a deviation of $\sigma=4$, and so
    a radius of $r=2\sigma=8$}
    \label{fig:normals-estimation}
\end{figure}

\paragraph{Experimental validation.}
Figure~\ref{fig:normals-estimation} shows an example
of our normals compared to normals computed with the integral invariant (II) estimator~\cite{Lachaud:2017-lnm}.
In order to compare quantitatively the different normal estimators, we compute the
root-mean-square error (RMSE) and the maximum error $E_{\max}$ of the
normals computed on a digital surface with respect to the normals computed
on the continuous surface (Figure~\ref{fig:errors-normals}). We use 4 different digital surfaces (ellipsoid,
goursat, rcube and sphere9). We see that the RMSE and $E_{\max}$ errors of our
estimator are convergent with respect to the grid resolution with a
rate of convergence of $h^{\frac{2}{3}}$ for the RMSE and $h^{\frac{1}{2}}$ for
$E_{\max}$.

\paragraph{Application to curvature estimation.}
In order to illustrate better that the visibility normals better
take into account the sharp features of a digital surface while
staying meaningfull in digitization of smooth parts, we compare
the curvature estimates induced by different discrete normal
estimators. We exploit the \emph{Corrected Normal Current (CNC)
    estimator}, whose theory is detailed in~\cite{lachaud:2022-dcg}
and which can estimate all kinds of curvatures given positions and
normals. It produces state-of-the-art curvature estimates on
polygonal surfaces~\cite{lachaud:2020-cgf}, point clouds~\cite{lachaud:2023-cgf}, and even outperforms Integral Invariant
curvature estimates~\cite{coeurjolly:2014-cviu} on digital
surfaces~\cite{lachaud:2022-dcg}. Its implementation is also
available in    \href{https://dgtal-team.github.io/doc-nightly/moduleCurvatureMeasures.html}{\textsc{DGtal}}.

Figure~\ref{fig:fig-curvatures} displays the results of CNC curvature
estimations on the digitization of a piecewise smooth shape
``Talking D20'', taken from
\href{https://ten-thousand-models.appspot.com/detail.html?file_id=1533028}{Thingi10K}.
We compare the differences obtained by just changing the discrete
normal estimator (available in the \textsc{DGtal} library):
(column II normals) Integral Invariant normals with $r=6$, (column
CTriv normals) convolved trivial normals with $r=6$, (column
Visibility normals) our proposed estimator (VN) with $\sigma=4$ and
maximal visibility distance $2\sigma$. These parameters were
chosen so that the respective computation windows of the different
estimators are approximately the same.

The II estimator is good along digitization of smooth or flat parts of
the dice, but presents curious artefacts near sharp features
induced by the hollowed out numbers: this is due to the nature of
II normal estimates, which computes a PCA of a ball centered on
the point of interest and may include points on the other side of
the saliencies.

The CTriv estimator only performs averaging of trivial normal
directions along the surface. It behaves much better than II near
features (although it tends to smooth curvatures) since it does
not use information from across the gap. However, some oscillations
of the curvatures are distinguishable especially along the flat
parts, and some curvature estimates are erroneous on smoother
parts (like the edge above 6 for $\kappa_2$ or several dice
vertices).

The VN estimator takes the best of the two previous approaches. It
remains precise and stable on smooth and flat regions, while
perfectly delineating sharp features and holes, whatever the kind
of estimated curvatures. Its drawback is the computation time,
which is $52s$ for a maximal visibility distance $8$, compared to
$2.3$s for II and $0.1s$ for CTriv. Note that the computation time
falls back to $33s$ for a maximal distance of $6$ (and same
$\sigma=4$) while the result is visually indistinguishable.

\newcommand{\MyZoom}[1]{%
    \begin{tikzpicture}[spy using outlines={circle,magnification=1.8,size=2cm,connect spies}]
    \node[inner sep=0pt] {\pgfimage[width=0.3\textwidth]{#1}};
    \spy[overlay,blue] on (0.4,0.2) in node at (-0.8,0.8);
    \end{tikzpicture}}

\begin{figure}
    \begin{center}
        \resizebox{!}{0.4\textheight}{%
            \begin{tabular}{|c||c|c|c|}
                \hline
                Curv. & II normals & CTriv normals & Our normals \\ \hline \hline
                \raisebox{18mm}{$\kappa_1$} &
                \MyZoom{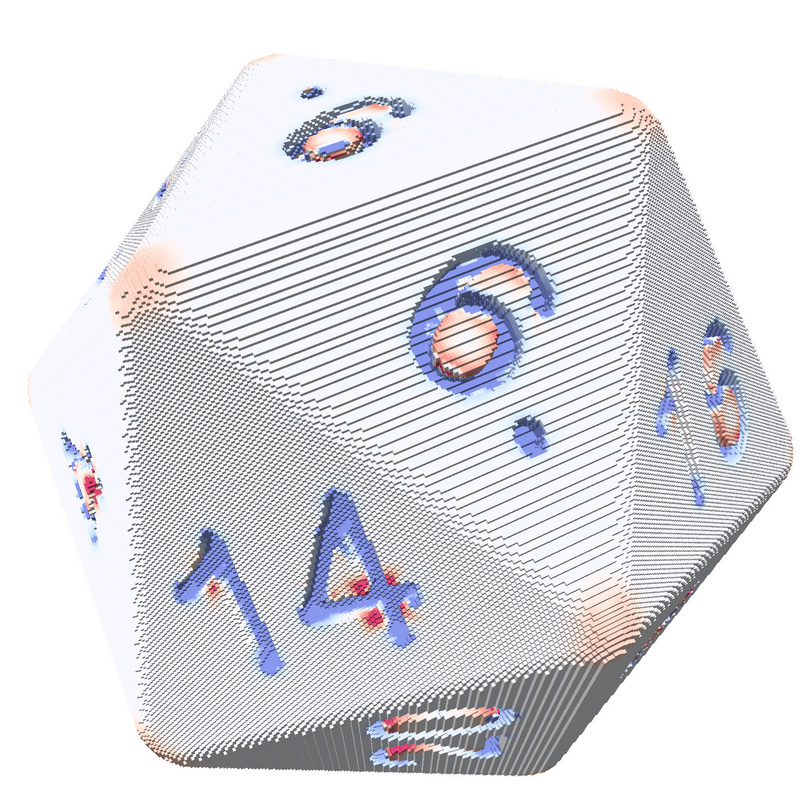} &
                \MyZoom{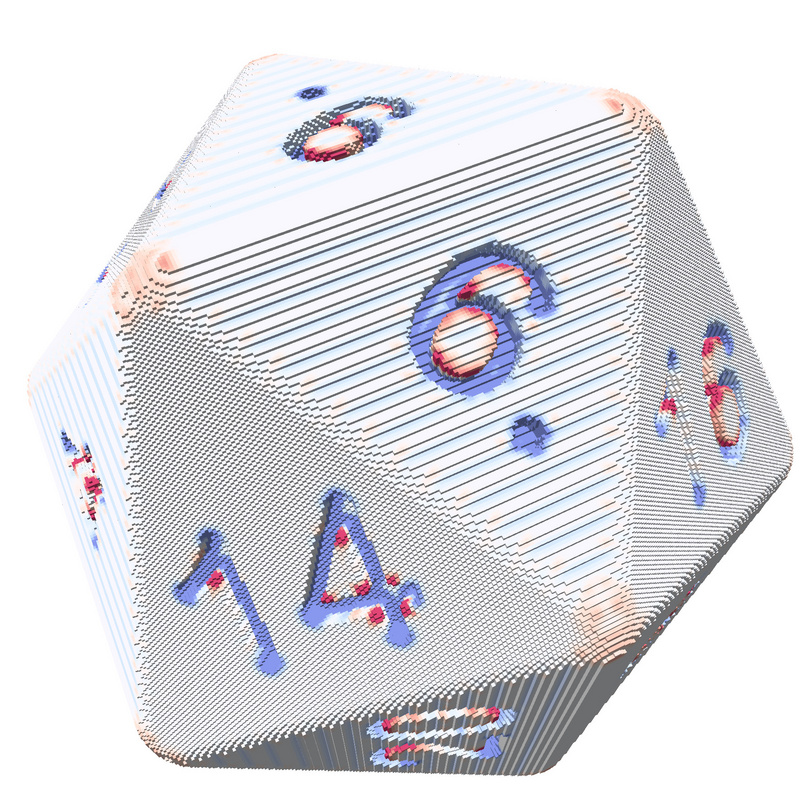}&
                \MyZoom{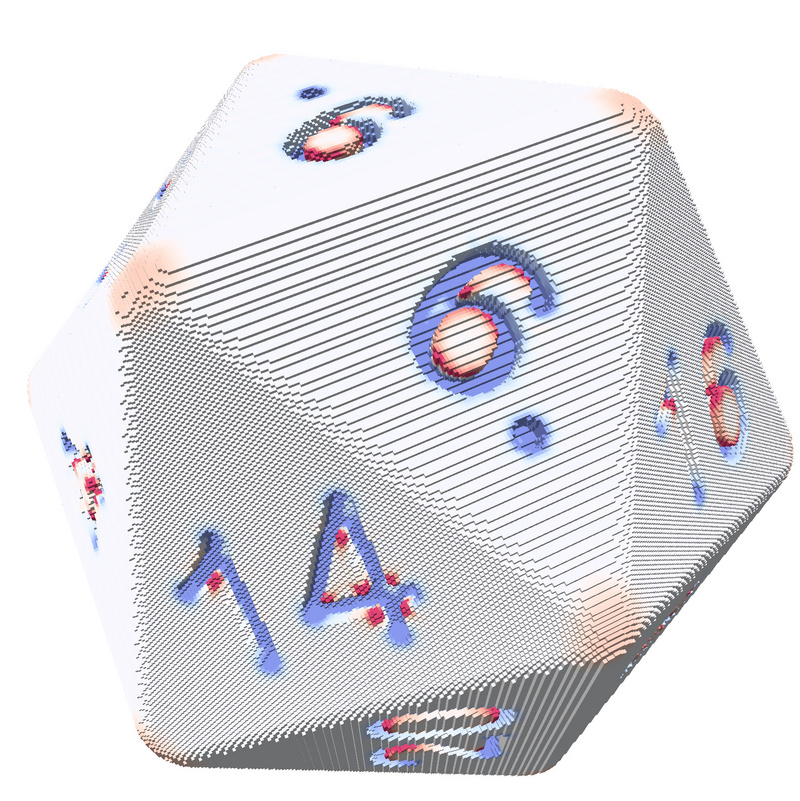}\\ \hline
                \raisebox{18mm}{$\kappa_2$} &
                \MyZoom{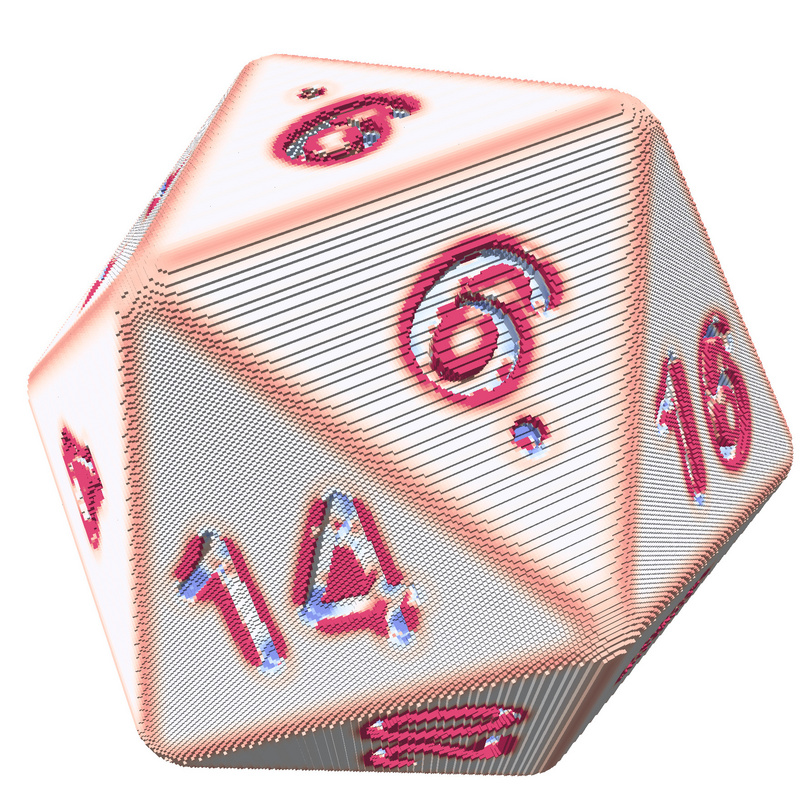} &
                \MyZoom{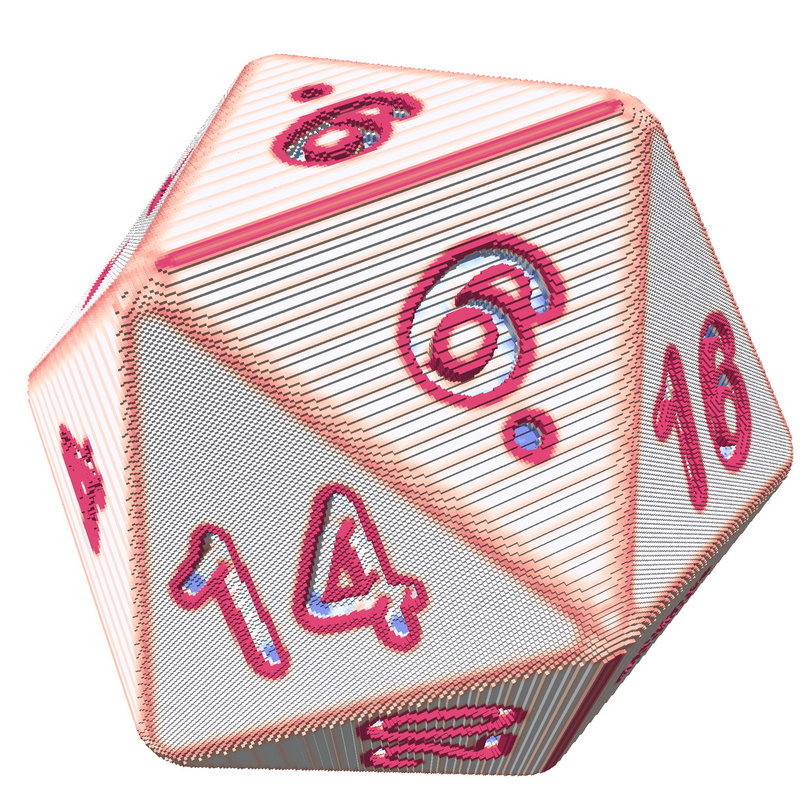}&
                \MyZoom{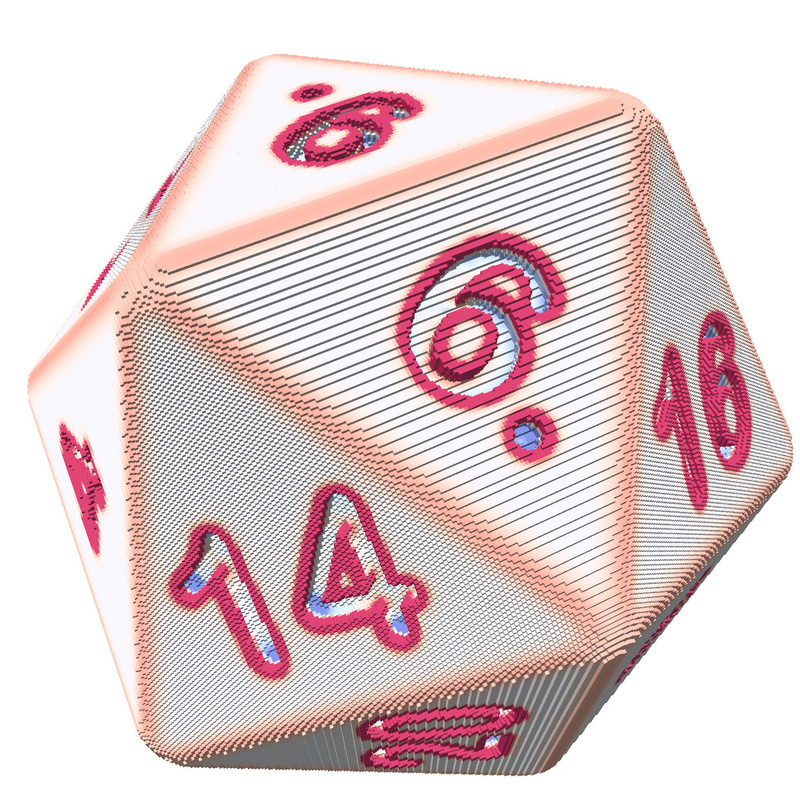}\\ \hline
                \raisebox{18mm}{$H$} &
                \MyZoom{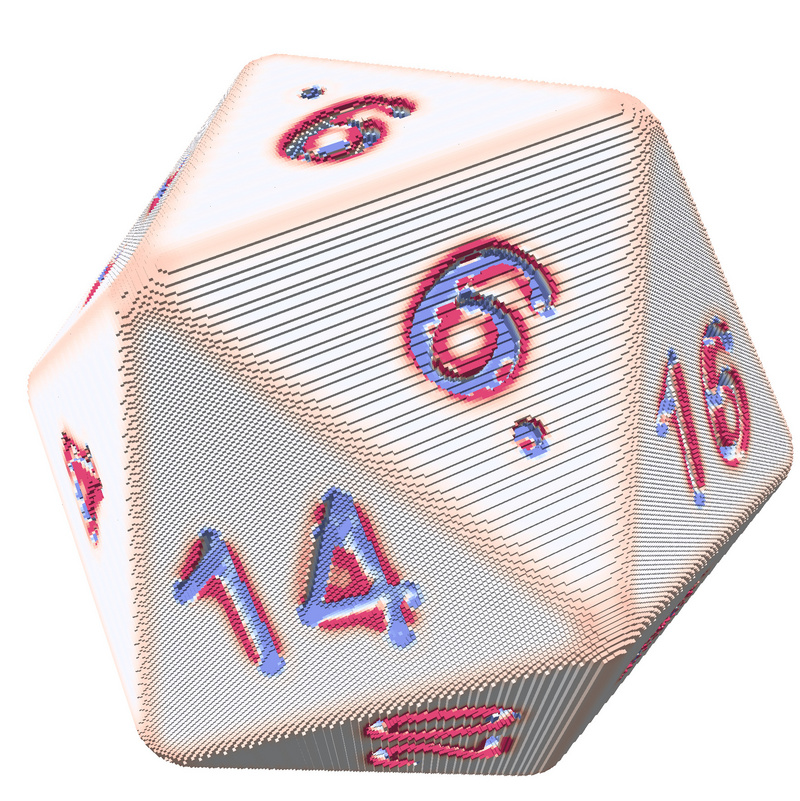} &
                \MyZoom{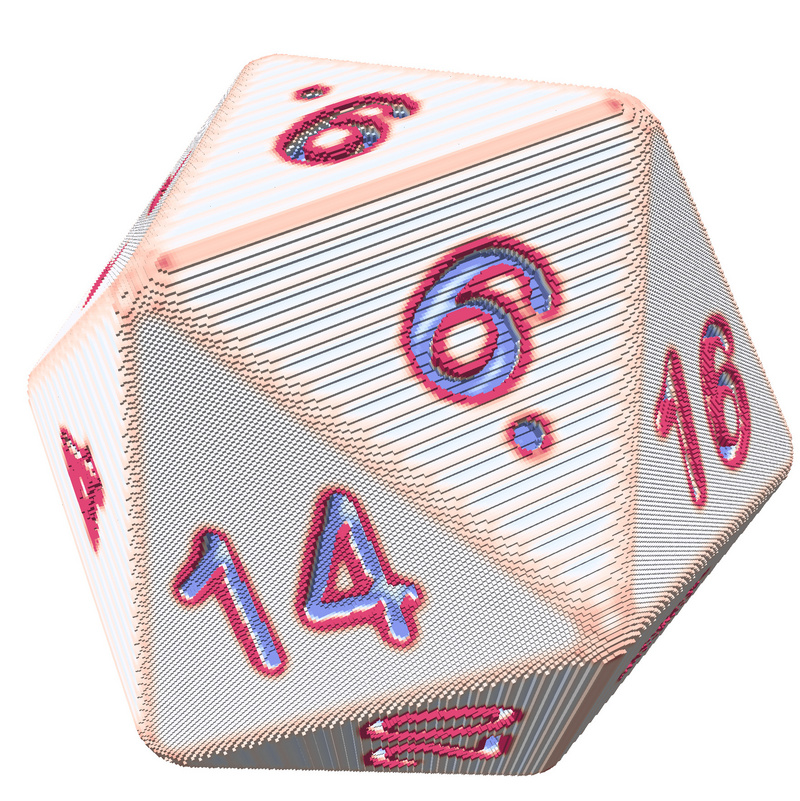}&
                \MyZoom{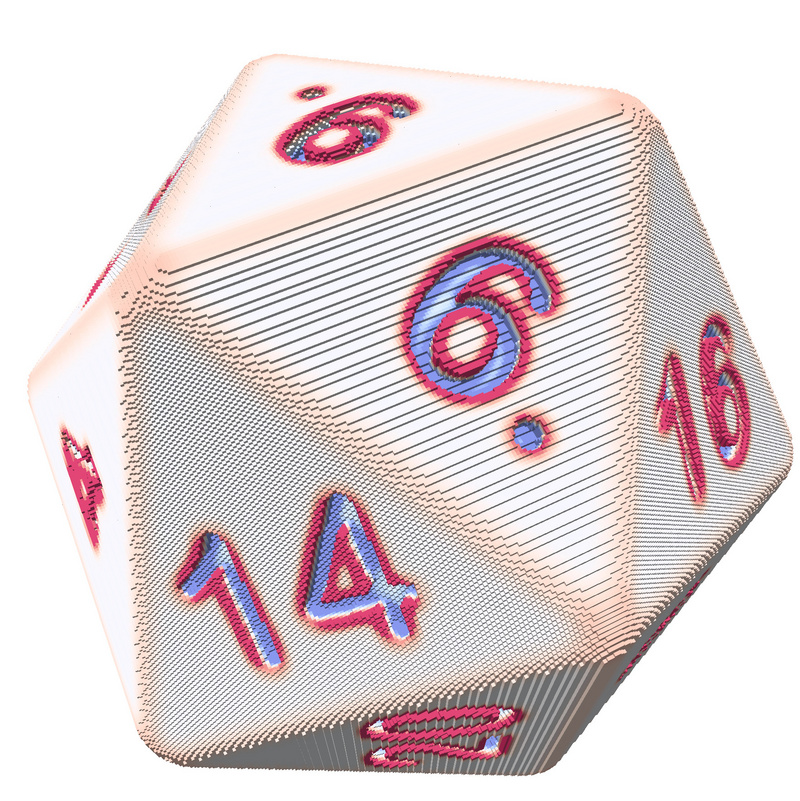}\\ \hline
                \raisebox{18mm}{$G$} &
                \MyZoom{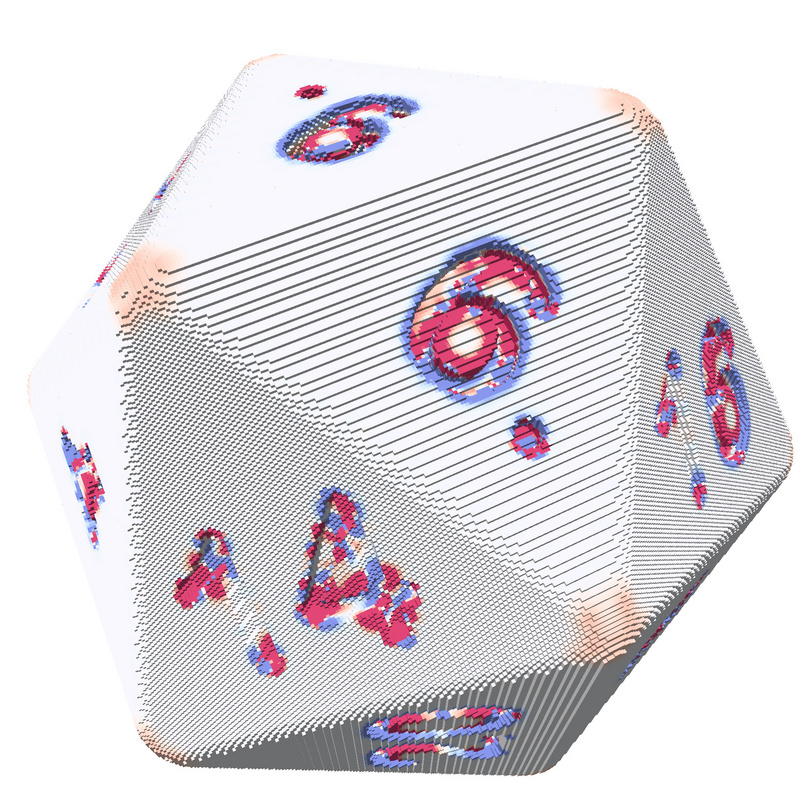} &
                \MyZoom{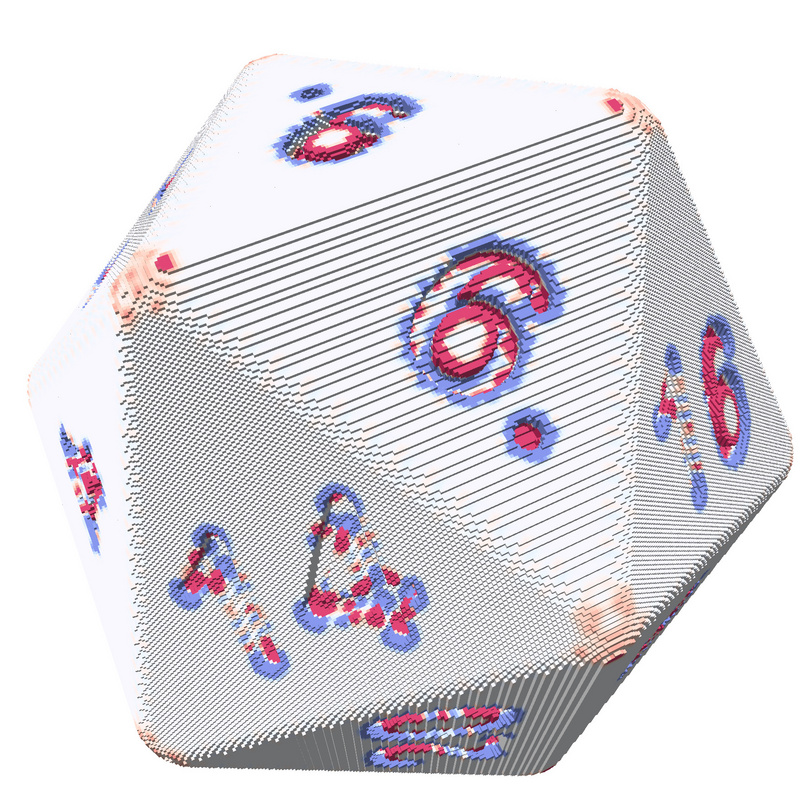}&
                \MyZoom{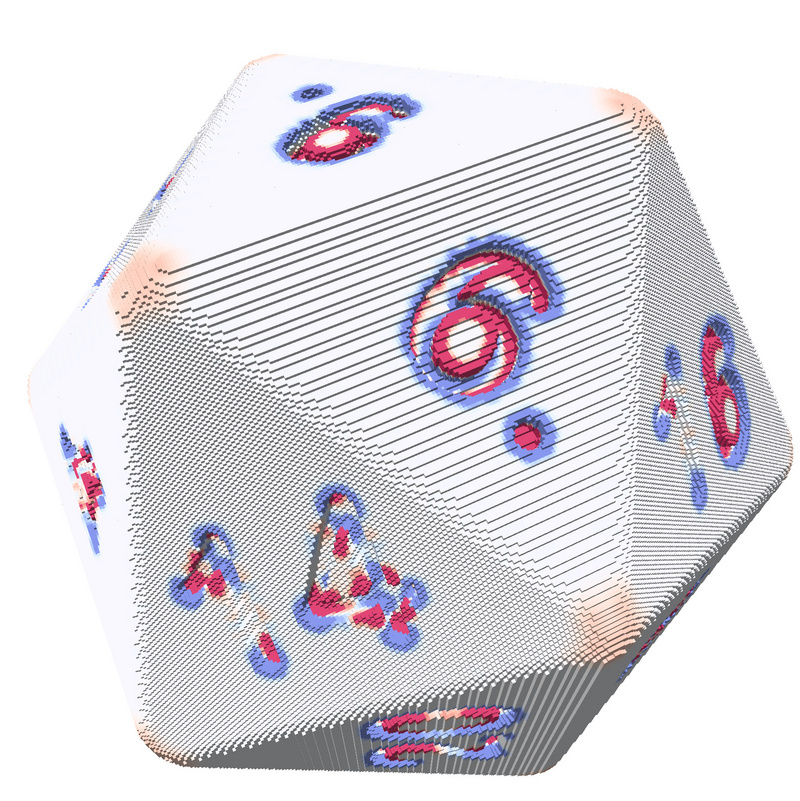}\\ \hline
            \end{tabular}%
        }
    \end{center}
    \caption{\label{fig:fig-curvatures}Estimation of curvatures using
    Corrected Normal Current estimators \cite{lachaud:2022-dcg}
    with a measure of radius $2$ for the shape ``Dice-20'' of
    Thingi10K database at resolution $256^3$. This estimator is
    parameterized by an input normal estimator: first column by
    ``Integral Invariant normals'' (radius $r=6$), second column
    by ``Convolved Trivial normals'' (radius $k=6$), third
    column by our presented normal estimator (deviation
        $\sigma=4$, radius $r=2\sigma=8$). Per row are displayed the
        estimated curvatures in order: first and second principal
        curvatures $\kappa_1$ and $\kappa_2$, mean curvature $H$,
        Gaussian curvature $G$. The range of curvature between deep
        blue and deep red is $\lbrack -0.1, 0.1 \rbrack$ with white
        color equal to $0$.}
\end{figure}

%% file: conclusion.tex
We have proposed a new algorithm that computes the visibility between
all pairs of points on a digital surface. It uses the specificity of a
structure to represent sets of lattice points or cells, a mapping from
projected coordinates to list of integer intervals. Compared to a
classical breadth-first approach to compute visibility, it provides
the exact visibility result, even when the set of visible points
from a source is disconnected. As shown by experiments the running
times are comparable with the breadth-first algorithm, and even faster
for small maximal visibility distance, which are typical when
processing standard 3d images up to $512^3$. We have used visibility
to define a discrete tangent estimator that respects salient features
while giving good approximations on digitization of smooth parts. The
obtained normal field is more suitable to curvature estimation than
the classical Integral Invariant normal estimator.

We have several ideas for possible future works. First, it is possible
to speed up the identification of non visibility by a coarse-to-fine
approach and a precomputation of a pyramid of covering cells. We would
like to explore if we can build such a pyramid variant of our
algorithm, considering the primitive vectors as a hierarchy. Second we
would like to prove the muligrid convergence of our normal estimator,
which is observable on experiments. Last, our algorithm is not limited
to pairwise visibility, but is more an algorithm of exact pattern
matching. We would like to explore new applications of pattern
identification along digital surfaces (like corner detection, locally
convex zones identification, etc.).

%% file: DGMM2025_arXiv.bbl
\begin{thebibliography}{10}
\providecommand{\url}[1]{\texttt{#1}}
\providecommand{\urlprefix}{URL }
\providecommand{\doi}[1]{https://doi.org/#1}

\bibitem{charrier:2011-iwcia}
Charrier, E., Lachaud, J.O.: Maximal planes and multiscale tangential cover of
  3d digital objects. In: Proc. Int. Workshop Combinatorial Image Analysis
  (IWCIA2011). LNCS, vol.~6636, pp. 132--143. Springer Berlin / Heidelberg
  (2011)

\bibitem{chica:2008-spm}
Chica, A.: Visibility-based feature extraction from discrete models. In:
  Proceedings of the 2008 ACM symposium on Solid and physical modeling. pp.
  347--352 (2008)

\bibitem{coeurjolly:2014-cviu}
Coeurjolly, D., Lachaud, J.O., Levallois, J.: Multigrid convergent principal
  curvature estimators in digital geometry. Computer Vision and Image
  Understanding  \textbf{129},  27--41 (2014)

\bibitem{coeurjolly:2004-prl}
Coeurjolly, D., Miguet, S., Tougne, L.: 2d and 3d visibility in discrete
  geometry: an application to discrete geodesic paths. Pattern Recognition
  Letters  \textbf{25}(5),  561--570 (2004)

\bibitem{Cuel:2014-dgci}
Cuel, L., Lachaud, J.O., Thibert, B.: Voronoi-based geometry estimator for 3d
  digital surfaces. In: Barcucci, E., Frosini, A., Rinaldi, S. (eds.) Proc.
  Int. Conf. on Discrete Geometry for Computer Imagery (DGCI'2014), Sienna,
  Italy. LNCS, vol.~8668, pp. 134--149. Springer International Publishing
  (2014). \doi{10.1007/978-3-319-09955-2_12}

\bibitem{durand:2002-tog}
Durand, F., Drettakis, G., Puech, C.: The 3d visibility complex. ACM
  Transactions on Graphics (TOG)  \textbf{21}(2),  176--206 (2002)

\bibitem{feschet:1999-dgci}
Feschet, F., Tougne, L.: Optimal time computation of the tangent of a discrete
  curve: Application to the curvature. In: Discrete Geometry for Computer
  Imagery: 8th International Conference, DGCI’99 Marne-la-Vall{\'e}e, France,
  March 17--19, 1999 Proceedings 8. pp. 31--40. Springer (1999)

\bibitem{fourey:2009-cg}
Fourey, S., Malgouyres, R.: Normals estimation for digital surfaces based on
  convolutions. Computers \& Graphics  \textbf{33}(1),  2--10 (2009)

\bibitem{ghosh:2007-book}
Ghosh, S.K.: Visibility algorithms in the plane. Cambridge university press
  (2007)

\bibitem{Lachaud:2017-lnm}
Lachaud, J.O., Coeurjolly, D., Levallois, J.: Robust and convergent curvature
  and normal estimators with digital integral invariants. In: Najman, L.,
  Romon, P. (eds.) Modern Approaches to Discrete Curvature, Lecture Notes in
  Mathematics, vol.~2184, pp. 293--348. Springer International Publishing, Cham
  (2017). \doi{10.1007/978-3-319-58002-9_9}

\bibitem{lachaud:2007-ivc}
Lachaud, J.O., Vialard, A., de~Vieilleville, F.: Fast, accurate and convergent
  tangent estimation on digital contours. Image and Vision Computing
  \textbf{25}(10),  1572--1587 (October 2007)

\bibitem{lachaud:2021-dgmm}
Lachaud, J.O.: An alternative definition for digital convexity. In: Lindblad,
  J., Malmberg, F., Sladoje, N. (eds.) Discrete Geometry and Mathematical
  Morphology - First International Joint Conference, DGMM 2021, Uppsala,
  Sweden, May 24-27, 2021, Proceedings. LNCS, vol. 12708, pp. 269--282.
  Springer (2021). \doi{10.1007/978-3-030-76657-3_19}

\bibitem{lachaud:2022-jmiv}
Lachaud, J.O.: An alternative definition for digital convexity. J. Math.
  Imaging Vis.  \textbf{64}(7),  718--735 (2022).
  \doi{10.1007/s10851-022-01076-0}

\bibitem{lachaud:2023-cgf}
Lachaud, J.O., Coeurjolly, D., Labart, C., and, P.R.: Lightweight curvature
  estimation on point clouds with randomized corrected curvature measures.
  Computer Graphics Forum  \textbf{42}(5),  e14910 (2023)

\bibitem{lachaud:2022-dcg}
Lachaud, J.O., Romon, P., Thibert, B.: Corrected curvature measures. Discret.
  Comput. Geom.  \textbf{68}(2),  477--524 (2022).
  \doi{10.1007/s00454-022-00399-4}

\bibitem{lachaud:2020-cgf}
Lachaud, J.O., Romon, P., Thibert, B., Coeurjolly, D.: Interpolated corrected
  curvature measures for polygonal surfaces. Computer Graphics Forum
  \textbf{39}(5),  41--54 (2020). \doi{10.1111/cgf.14067}

\bibitem{mareche:2024-ispr}
Mar{\^e}ch{\'e}, A., Debled-Rennesson, I., Feschet, F., Ngo, P.: A
  parameter-free normal estimator on digital surfaces. In: International
  Conference on Intelligent Systems and Pattern Recognition. pp. 130--142.
  Springer (2024)

\bibitem{nguyen:2011-pr}
Nguyen, T.P., Debled-Rennesson, I.: A discrete geometry approach for dominant
  point detection. Pattern Recognition  \textbf{44}(1),  32--44 (2011)

\bibitem{nirenstein:2002-ewr}
Nirenstein, S., Blake, E., Gain, J.: Exact from-region visibility culling. In:
  Proceedings of 13th Eurographics Workshop on Rendering (2002)

\bibitem{orourke:2017-book}
O’Rourke, J.: Visibility. In: Handbook of discrete and computational
  geometry, pp. 875--896. Chapman and Hall/CRC (2017)

\bibitem{soille:1994-prl}
Soille, P.: Generalized geodesy via geodesic time. Pattern Recognition Letters
  \textbf{15}(12),  1235--1240 (1994)

\bibitem{soille1999morphological}
Soille, P., et~al.: Morphological image analysis: principles and applications,
  vol.~2. Springer (1999)

\end{thebibliography}
